\documentclass[12pt]{article}
\usepackage{apacite}
\usepackage{amssymb,amsfonts,amsmath}
\usepackage{times}
\usepackage{graphicx}
\usepackage{tikz}
\usetikzlibrary{bayesnet}
\usepackage{color}
\usepackage{float}
\usepackage{multirow}
\usepackage[authoryear]{natbib}
\usepackage{rotating}
\usepackage{bbm}
\usepackage{pgf}
\usepackage{caption}
\usepackage{import}
\usepackage{latexsym}
\allowdisplaybreaks[1]
\newcommand{\captionfonts}{\normalsize}

\makeatletter
\long\def\@makecaption#1#2{%
  \vskip\abovecaptionskip
  \sbox\@tempboxa{{\captionfonts #1: #2}}%
  \ifdim \wd\@tempboxa >\hsize
    {\captionfonts #1: #2\par}
  \else
    \hbox to\hsize{\hfil\box\@tempboxa\hfil}%
  \fi
  \vskip\belowcaptionskip}
\makeatother

\newcommand{\w}{\mathbf{w}}
\newcommand{\x}{\mathbf{x}}
\newcommand{\z}{\mathbf{z}}
\newcommand{\ess}{\mathbf{s}}
\newcommand{\vv}{\mathbf{v}}
\newcommand{\A}{\mathbf{A}}

\newcommand{\R}{{\mathbb{R}}}

\newtheorem{thm}{Theorem}

\newtheorem{defn}[thm]{Definition}

\begin{document}
\hspace{13.9cm}1

\ \vspace{20mm}\\

\begin{center}
{\LARGE Estimating a Separably-Markov Random Field (SMuRF) from Binary Observations}
\end{center}

\ \\
{\bf \large Yingzhuo Zhang$^{\displaystyle 1}$, Noa Malem-Shinitski$^{\displaystyle 2}$, Stephen A Allsop$^{\displaystyle 3}$, Kay Tye$^{\displaystyle 3}$ and Demba Ba$^{\displaystyle 1}$}\\
{$^{\displaystyle 1}$ Harvard University, John A. Paulson School of Engineering and Applied Sciences.}\\
{$^{\displaystyle 2}$ Technische Universit\"{a}t Berlin.}\\
{$^{\displaystyle 3}$ Massachusetts Institute of Technology, Department of Brain and Cognitive Sciences.}\\
%

\noindent {\bf Keywords:} Spike rasters, Dynamics, Random field, Bayesian Estimation, associative learning.

\thispagestyle{empty}
\markboth{}{NC instructions}
\ \vspace{-0mm}\\
%
\begin{center} {\bf Abstract} \end{center}
A fundamental problem in neuroscience is to characterize the dynamics of spiking from the neurons in a circuit that is involved in learning about a stimulus or a contingency. A key limitation of current methods to analyze neural spiking data is the need to collapse neural activity over time or trials, which may cause the loss of information pertinent to understanding the function of a neuron or circuit. We introduce a new method that can determine not only the trial-to-trial dynamics that accompany the learning of a contingency by a neuron, but also the latency of this learning with respect to the onset of a conditioned stimulus. The backbone of the method is a separable two-dimensional (2D) random field (RF) model of neural spike rasters, in which the joint conditional intensity function of a neuron over time and trials depends on two latent Markovian state sequences that evolve separately but in parallel. 
Classical tools to estimate state-space models cannot be applied readily to our 2D separable RF model. We develop efficient statistical and computational tools to estimate the parameters of the separable 2D RF model. We apply these to data collected from neurons in the pre-frontal cortex (PFC) in an experiment designed to characterize the neural underpinnings of the associative learning of fear in mice.
Overall, the separable 2D RF model provides a detailed, interpretable, characterization of the dynamics of neural spiking that accompany the learning of a contingency. 



\section{Introduction}

A fundamental problem in the analysis of electrophysiological data from neuroscience experiments is to determine the trial, and time within said trial, when a neuron or circuit first exhibits a conditioned response to a stimulus. This is a challenging problem because neural spike rasters resulting from such experiments can exhibit variability both within a given trial and across trials~\citep{Czanner:08}. Fear conditioning experiments~\citep{allsop2014optogenetic} are a prime example of a scenario when this situation arises: a neutral stimulus, present across all trials of an experiment, gives rises to stereotypical within-trial spiking dynamics, while the associated aversive stimulus leads to changes in spiking dynamics across a subset of the trials.


State-of-the-art methods for analyzing neural spike rasters fall primarily within two classes. The most pervasive class of such methods neglect the inherent two-dimensional nature of neural spike rasters by aggregating the raster data either across time or trials, and subsequently applying techniques applicable to one-dimensional signals~\citep{smith2003estimating,Zammit:2012,Yuan:2012,scott2012fully}. In contrast to these one-dimensional methods, two-dimensional methods model both the within and cross-trial dynamics of neural spiking~\citep{Czanner:08,rad2010efficient}. Within the class of one-dimensional methods, the past decade has seen a growing interest in approaches based on state-space models of neural spiking activity. These approaches treat neural spiking data as realizations of a stochastic point process whose conditional intensity function obeys a stochastic smoothness constraint in the form of a Markov process followed by a nonlinearity. The main challenge is to estimate the parameters of the model, and various solutions have been proposed towards this end~\citep{smith2003estimating,Zammit:2012,Yuan:2012,scott2012fully}. The main drawback of one-dimensional approaches applied to the analysis of neural spike rasters is the need, preceding analysis, for aggregation across one of the dimensions. Among one-dimensional methods, non-parametric methods based on rank tests (e.g. Wilcoxon rank sum test) have been the most popular, primarily due to their ease of application. In addition to the need to collapse neural activity of time or trials, two common pitfalls of non-parametric methods are their reliance on large sample assumptions to justify comparing neural spiking rates, and the need to correct for multiple comparisons. {For instance, tests that rely on estimates of the neural spiking rate based on empirical averages are hard to justify when it is of interest to characterize the dynamics of neural spiking at the millisecond time scale. Consider a neural spike raster for which it is of interest to assess differences in instantaneous spiking rates between distinct time/trial pair. At the millisecond time scale, there would only be one observation per time/trial pair, violating the large sample assumptions that such non-parametric methods rely upon.} To the best of our knowledge, the work of~\citep{Czanner:08} remains the most successful attempt to characterize simultaneously the within and cross-trial dynamics of neural spiking. This approach uses a state-space model of the cross-trial dynamics, in conjunction with a \emph{parametric} model of the within-trial dynamics. The use of a parametric model for the within-trial dynamics is convenient because it enables the estimation of the model parameters by Expectation-Maximization (EM), using a combination of point-process filtering and smoothing in the E-step (to fill-in the missing cross-trial effect), and an M-step for the within-trial parameters that resembles a GLM~\citep{Truc:05}. The main drawbacks of this approach are, on the one hand, the high-dimensionality of the state-space model that captures the cross-trial dynamics, and on the other hand the lack of a simple interpretation, 
as in the one-dimensional models~\citep{smith2003estimating,Zammit:2012,Yuan:2012}, for the state sequence. 
{Lastly, a two-dimensional approach based on Gaussian processes was proposed in~\citep{rad2010efficient}. One advantage of this approach, which is based on a Gaussian process prior of the neural spiking rate surface, is its ability to model the interaction between the two dimensions through the use of a two dimensional kernel. As is common with kernel methods, it does not scale well to multiple dimensions.} 

We propose a two-dimensional (2D) random field (RF) model of neural spike rasters--termed Separably-Markov Random Field (SMuRF)--in which the joint conditional intensity function of a neuron over time and trials depends on two latent Markovian state sequences that evolve separately but in parallel. Conventional methods for estimating state-space models from binary observations~\citep{smith2003estimating,Zammit:2012,Yuan:2012} are not applicable to SMuRF. We derive a Monte Carlo Expectation-Maximization algorithm to maximize the marginal likelihood of observed data under the SMuRF model. In the E-step, we leverage the Polya-Gamma~\citep{polson2013bayesian} representation of Bernoulli random variables to generate samples from the joint posterior distribution of the state sequences by Gibbs sampling. A similar strategy was adopted in~\citep{scott2012fully} for a one-dimensional state-space model. The sampler uses a highly efficient forward-filtering backward-sampling algorithm for which the forward step can be implemented \emph{exactly} and elegantly as Kalman filter, while the backward step uses Bayes' rule to correct the filter samples. The SMuRF model obviates the need for aggregation across either time or trials, and yields a \emph{low-dimensional} 2D characterization of neural spike rasters that is interpretable 
in the sense that the posterior of the two state sequences capture the variability within and across trials respectively. Moreover, being model-based, the SMuRF model, unlike non-parametric methods, yields a characterization of the joint posterior (over all trials and time within a trial) distribution of the instantaneous rate of spiking, thus allowing us to precisely determine the dynamics of neural spiking that accompany the learning of a contingency. To demonstrate this, we  apply the model to data collected from neurons in the pre-frontal cortex (PFC) in an experiment designed to characterize the neural underpinnings of the associative learning of fear in mice. We find that the trial at which the cortical neurons begin to exhibit a conditioned response to the auditory conditioned stimulus  
is robust across cells, occurring $3$ to $4$ trials into the conditioning period. We also find that the time with respect to conditioned stimulus onset when we observe a significant change in neural spiking compared to baseline activity varies significantly from cell to cell, occurring between $20$ to $600$ ms after conditioned stimulus onset. These findings are likely reflective of the variability in synaptic strength and connectivity that accompany learning, as well as the location of the neurons in the population.

The rest of our treatment begins in Section~\ref{sec:model} where we motivate the SMuRF model, define it and introduce our notation. In Section~\ref{sec:ml}, we present the Monte-Carlo EM algorithm for parameter estimation in the SMuRF model, as well as our process for inferring the dynamics of neural spiking that accompany the learning of a contingency by a neuron. The reader may find derivations relevant to this section in the Appendix. We present an application to the cortical data in Section~\ref{sec:res}, and conclude in Section~\ref{sec:disco}.

\section{Notation and SMuRF model}
\label{sec:model}

We begin this section with a continuous-time point-process formalism of a neural spike raster, characterized by a trial-dependent conditional intensity function (CIF). Then, we introduce the SMuRF model, a model for the discrete-time version of the CIF.

\subsection{Continuous-time point-process observation model}

We consider an experiment that consists of $R$ successive trials. During each trial, we record the activity of a neuronal spiking unit. We assume, without loss of generality, that the duration of the observation interval during each trial is $(0,T]$. For trial $r$, $r=1,\cdots,R$, let the sequence {{$0 < t_{r,s} < \cdots < t_{r,{S_r}} < T$}} correspond to the times of occurrence of events from the neuronal unit, that is to say the times when the membrane potential of the neuron crosses a given threshold. We assume that {{$\{t_{r,s}\}_{s=1}^{S_r}$}} is the realization in $(0,T] $ of a stochastic point-process with counting process $N_r(t) = \int_{0}^t dN_r(u)$, where $dN_r(t)$ is the indicator function in $(0,T]$ of {{$\{t_{r,s}\}_{s=1}^{S_r}$}}. A point-process is fully characterized by its CIF. Let $\lambda_r(t|H_t)$ denote the trial-dependent CIF of $dN_r(t)$ defined as
\begin{equation}
	\lambda_r(t|H_t) = \lim_{\Delta \rightarrow 0} \frac{P[N_r(t+\Delta)-N_r(t)=1|H_t]}{\Delta},
\end{equation}
where $H_t$ is the history of the point process up to time $t$.

We denote by {{$\{\Delta N_{k,r}\}_{k=1,r=1}^{K,R}$}}, the discrete-time process obtained by sampling $dN_r(t)$ at a resolution of $\Delta$, $K = \left \lfloor \frac{T}{\Delta} \right \rfloor$. Let $\{\lambda_{k,r}\}_{k=1,r=1}^{K,R}$ denote the discrete-time, trial-dependent, CIF of the neuron.

\subsection{Separably-Markov Random Field (SMuRF) model of within and cross-trial neural spiking dynamics}

	\begin{defn}
	{Let $\{y_{k,r}\}_{k=1,r=1}^{K,R} \in \R^{K \times R}$ be a collection of random variables. We say that this collection is a separable random field if $\exists$ $\mathbf{x} \in \R^{K}, \mathbf{z} \in \R^{R}$ s.t. $\forall k,r$ $\exists$ unique $(x_k,z_r) \in \mathbf{x} \times \mathbf{z}$ s.t. $y_{k,r}|(\mathbf{x},\mathbf{z}) \sim f(x_k,z_r)$.}
	\\
	\vspace{0.15in}
	If in addition $\mathbf{x}$ and $\mathbf{z}$ are Markov processes, we say that  $\{y_{k,r}\}_{k=1,r=1}^{K,R}$ is a separably-Markov random field or ``SMuRF".
	\end{defn}

{A 2D random field $\{y_{k,r}\}_{k=1,r=1}^{K,R} \in \R^{K \times R}$ is a collection of random variables indexed over a subset of $\mathbb{N}^+ \times \mathbb{N}^+$. We call this collection a separable field if there exists latent random vectors $\x$ and $\z$ (each indexed over a subset of $\mathbb{N}^+$) such that $\{y_{k,r}\}_{k=1,r=1}^{K,R}$ are independent \emph{conditioned} on $\x$ and $\z$ and \emph{only} a function of the outer product between $\x$ and $\z$. If, in addition, $\x$ and $\z$ are Markov, we say that the field is a SMuRF. Intuitively, a separable random field is a random field that admits a stochastic rank-one decomposition.}

We propose the following SMuRF model of the discrete-time, trial-dependent, CIF $\{\lambda_{k,r}\}_{k=1,r=1}^{K,R}$ of a neuronal spiking unit
{{\begin{equation}\left\{
\begin{array}{ll}
	  x_k = \rho_x x_{k-1} + \alpha_x u_{x,k} + \epsilon_k , \epsilon_k \sim \mathcal{N}(0, \sigma^2_\epsilon) \\
      z_r = \rho_z z_{r-1} + \alpha_z u_{z,k}+\delta_r , \delta_r \sim \mathcal{N}(0,\sigma^2_\delta) \\
      \text{log } \frac{\lambda_{k,r}\Delta}{1-\lambda_{k,r}\Delta} = x_k + z_r \\
      \Delta N_{k,r} | x_k, z_r \sim \text{Bernoulli}(\lambda_{k,r}\Delta) \\
\end{array}
\right.
\label{eqn:smurf_model}
\end{equation}}}
\noindent By construction, this is a SMuRF of the trial-dependent CIF of a neuron. {{$u_{x,k}$ and $u_{z,k}$ are indicator functions of presence of cue.}} {To provide some intuition, \emph{if we assume $x_k + z_r$ is small}, then the SMuRF model approximates the trial-dependent CIF as $\lambda_{k,r}\Delta \approx \text{e}^{z_r}\cdot\text{e}^{x_k}$, that is, as the product of a within-trial component $\text{e}^{x_k}$ in units of Hz (spikes/s) and a unitless quantity $\text{e}^{z_r}$. For a given trial $r$, $\text{e}^{z_r}$ represents the excess spiking rate above what can be expected from the within-trial component at that trial, which we call the cross-trial component of the CIF. The within and cross-trial components from the SMuRF model are functions of two independent state sequences, $(x_k)_{k=1}^K$ and $(z_r)_{r=1}^R$, that evolve smoothly according to a first-order stochastic difference equation.  {The parameters $\rho_x$, $\alpha_x$, $\sigma^2_\epsilon$, $\rho_z$, $\alpha_z$ and $\sigma^2_\delta$, which govern the smoothness of $(x_k)_{k=1}^K$ and $(z_r)_{r=1}^R$, must be estimated from the raster data.}
\\
\noindent \underline{\textbf{Remark 1}}: We note that, in its generality, our model does not assume that $\lambda_{k,r}\Delta = \text{e}^{z_r}\cdot\text{e}^{x_k}$. In our model, $\lambda_{k,r}\Delta = \frac{\text{e}^{x_k + z_r}}{1+\text{e}^{x_k + z_r}}$. The approximation $\lambda_{k,r}\Delta \approx \text{e}^{z_r}\cdot\text{e}^{x_k}$ holds for a neuron with small neural spiking rate~\citep{Truc:05}.} 

\begin{figure}
  \begin{center}
  	\includegraphics[scale=1]{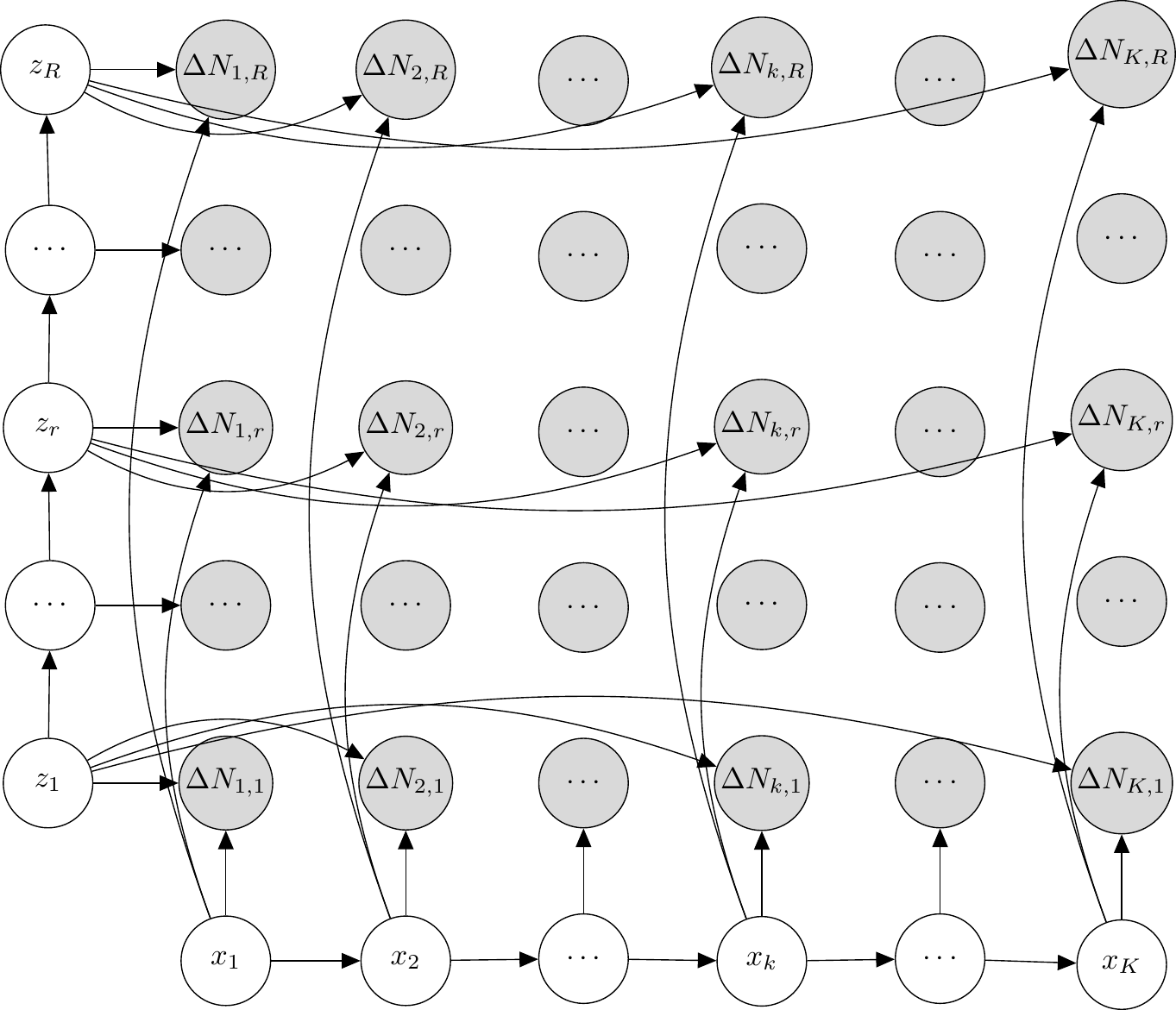}
    \end{center}
    \caption{{Representation of the SMuRF model as a Bayesian network. The SMuRF model approximates the trial-dependent CIF the product of a within-trial component in units of Hz
(spikes/s) and a unitless quantity. The within and cross-trial components are
functions of two independent state sequences, $(x_k)_{k=1}^K$ and $(z_r)_{r=1}^R$, that evolve smoothly,
each according to a first-order stochastic difference equation. Observations from the raster are Bernoulli random variables whose probability of of occurrence is a nonlinear function of the sum of the two sequences.}}
    \label{fig:bysnt}
\end{figure}

Figure~\ref{fig:bysnt} shows a graphical representation of the SMuRF model as a Bayesian network. {{It is not mathematically possible to rewrite the state equations from the SMuRF model in standard state-space form \emph{without} increasing significantly the dimension of the state space. We give a sketch of an argument as to why in the Appendix}}. {Therefore, in an Expectation-Maximization (EM) algorithm for parameter estimation, one cannot simply apply classical (approximate) binary filtering and smoothing in the E-step~\citep{smith2003estimating}}. We derive a Monte-Carlo Expectation-Maximization algorithm to maximize the likelihood of observed data under the SMuRF model, with respect to the parameter vector {{$\theta = (\rho_x,\alpha_x,\sigma^2_\epsilon,\rho_z,\alpha_z,\sigma^2_\delta)$}}.

\section{Parameter Estimation in the SMuRF by Maximum Likelihood}
\label{sec:ml}

\subsection{Maximum Likelihood Estimation by Expectation-Maximization}

Let $\x = (x_1,\cdots,x_K)^{\text{T}}$, $\mathbf{z} = (z_1,\cdots,z_R)^{\text{T}}$, and 
	$\mathbf{\Delta N} = \{\Delta N_{k,r}\}_{k=1,r=1}^{K,R}$. The goal is to maximize, with respect to $\theta$, the likelihood $L(\theta|\mathbf{\Delta N})$ of the SMuRF model
\begin{equation}
	L(\theta|\mathbf{\Delta N}) = \text{log } p(\mathbf{\Delta N}; \theta) = \text{log } \int_{\x,\z} p(\mathbf{\Delta N},\x,\z; \theta) d\x d\z.
	\label{eq:ll}
\end{equation}

\noindent This is a challenging problem because of the high-dimensional integral that must be carried out in Equation~\ref{eq:ll}. We propose to maximize the likelihood by EM.
\\
\noindent \underline{\textbf{Remark 2}}: For the moment, we treat $\x$ and $\z$ as missing data; in the sequel, we will augment the model with additional missing data that will simplify the EM algorithm.
\\
Given a candidate solution $\theta^{(\ell)}$, EM~\citep{dempster1977maximum} maximizes $L(\theta|\mathbf{\Delta N})$ by building a sequence of successive approximations $\mathcal{Q}(\theta|\theta^{(\ell)})$ of $L(\theta|\mathbf{\Delta N})$ (the so-called E-step) such that maximizing these approximations, which in general is simpler than directly maximizing $L(\theta|\mathbf{\Delta N})$, is guaranteed to not decrease $L(\theta|\mathbf{\Delta N})$. That is, each iteration of EM generates a new candidate solution $\theta^{(\ell+1)}$ such that $L(\theta^{(\ell+1)}|\mathbf{\Delta N}) \geq L(\theta^{(\ell)}|\mathbf{\Delta N})$. By iterating this process, EM generates a sequence of iterates $\{\theta^{(\ell)}\}_{\ell=1}^\infty$ that, under regularity conditions, converge to a local optimum of $L(\theta|\mathbf{\Delta N})$~\citep{dempster1977maximum}.

In the context of the SMuRF model, the key challenge of EM is to compute  $\mathcal{Q}(\theta|\theta^{(\ell)})$ defined as
\begin{equation}
	\mathcal{Q}(\theta|\theta^{(\ell)}) = \mathbb{E}_{\mathbf{x},\mathbf{z}}\left[\text{log } p(\mathbf{\Delta N},\mathbf{x},\mathbf{z};\theta)| \mathbf{\Delta N},\theta^{(\ell)}\right],
\end{equation}
the expected value of the \emph{complete-data likelihood} with respect to the \emph{joint} posterior distribution of the missing data $(\x,\z)$ conditioned on the observed data $\mathbf{\Delta N}$ and the candidate solution $\theta^{(\ell)}$. This expectation is not tractable, i.e. it cannot be computed in closed-form. The intractability stems not only from the lack of conjugacy between the Bernoulli observation model and our Gaussian priors--also an issue for one-dimensional models~\citep{smith2003estimating}--but also because, as mentioned previously, the SMuRF model cannot be reduced to a standard state-space model.  We propose to approximate the required expectations using Markov-Chain Monte-Carlo (MCMC) samples from $p(\x,\z|\mathbf{\Delta N};\theta^{(\ell)})$. In particular, we will use Gibbs sampling~\citep{casella1992explaining}, a Monte-Carlo technique, to generate samples from a distribution by sampling from its so called full conditionals (conditional distribution of one variable given all others), thus generating a Markov chain that, under regularity conditions, can be shown to converge to a sample from the desired distribution. Gibbs sampling is attractive in cases where sampling from the full-conditionals is simple. However, it is prone to the drawbacks of MCMC methods, such as poor mixing and slow convergence, particularly if one is not careful in selecting the full-conditionals from which to generate samples from. Two observations are in order, that will lead to the derivation of an elegant block Gibbs sampler with attractive properties
\begin{itemize}
\item Conditioned on $\z$, the joint distribution, $p(\mathbf{\Delta N},\x|\z;\theta)$, of $\x$ and $\mathbf{\Delta N}$ is equivalent to the joint distribution from a \emph{one-dimensional} state-space model with binary observations~\citep{smith2003estimating}. By symmetry, this is also true for $p(\mathbf{\Delta N},\z|\x;\theta)$. This readily motivates a block Gibbs sampler that alternates between sampling from $\x|\mathbf{\Delta N};\theta$ and $\z | \mathbf{\Delta N};\theta$. This leaves us with one challenge: how to obtain samples from the posterior distribution of the state in a one-dimensional state-space model with Bernoulli (more generally binomial) observations?
\item We introduce a new collection of i.i.d., Polya-Gamma distributed~\citep{polson2013bayesian} random variables $\mathbf{w} = \{w_{k,r}\}_{k=1,r=1}^{K,R}$, such that sampling from $\x|\mathbf{\Delta N},\w;\theta$ is equivalent to sampling from the posterior of the state in a \emph{linear Gaussian} state-space model (we will prove this in the Appendix) using a forward-filtering backward-sampling algorithm~\citep{fruhwirth1994data}. Moreover, it has been shown that the Gibbs sampler based on this Polya-Gamma augmentation scheme~\citep{choi2013polya} is uniformly ergodic and possesses superior mixing properties to alternate data-augmentation scheme for logit-based models~\citep{polson2013bayesian,choi2013polya}. {The intuition behind the introduction of the Polya-Gamma random variables is the following: they are missing data that, if we could observe, would make the Bernoulli observations Gaussian. Stated otherwise, the Polya-Gamma random variables are scale variables in a Gaussian scale mixture~\citep{andrews1974scale} representation of Bernoulli random variables.}
\end{itemize}

{
\noindent \underline{\textbf{Remark 3}}: The random vector $\w$ in the preceding bullet point is the vector additional missing data alluded to in \textbf{Remark 2}.
}

\noindent Together, these two observations form the basis of an efficient block-Gibbs sampler we use for maximum-likelihood estimation of the parameters from the SMuRF model by Monte-Carlo EM, also referred to as empirical Bayes~\citep{casella2001empirical}. We introduce the basic ideas behind PG augmentation and its utility in Bayesian estimation for logit-based models. In the Appendix, we provide detailed derivations for the PG sampler adapted to the SMuRF model.

\subsection{Polya-Gamma augmentation and sampling in one dimension}

{

Let $\Delta N \in \{0,1\}$ and $X \in \R$ and suppose that, conditioned on $X = x$, $\Delta N$ is Bernoulli with mean $\frac{\text{e}^x}{1+\text{e}^x}$, i.e.
\begin{equation}
	p(\Delta N|x) = \frac{(\text{e}^x)^{\Delta N}}{1+\text{e}^x}
\end{equation}

\noindent We begin with a definition of Polya-Gamma (PG) random variables, followed by a PG augmentation scheme for the Bernoulli/binomial likelihood. We will see that the augmentation scheme leads to an attractive form for the posterior of $x$ given the observation $\Delta N$ and the augmented variable. Finally, we will see that the posterior of the augmented variable itself follows a PG distribution. Our treatment follows closely that of~\citep{choi2013polya}.

\noindent{\textbf{Definition of Polya-Gamma random variables}}: Let $\{E_m\}_{m=1}^\infty$ be a sequence of i.i.d. exponential random variable with parameter equal to 1. The random variable
\begin{equation}
	W \overset{d}{=} \frac{2}{\pi^2} \sum_{m=1}^\infty \frac{E_m}{(2m-1)^2}
\end{equation}
\noindent follows a PG(1,0) distribution, where $\overset{d}{=}$ denotes equality in distribution. The moment generating function of $W$ is
\begin{equation}
	\mathbb{E}[\text{e}^{-tW}] = \text{cosh}^{-1}\left(\frac{\sqrt{t}}{2}\right).
\end{equation}
\noindent An expression for its density $p_W(w)$, expressed as an infinite sum, can be found in~\citep{choi2013polya} and~\citep{polson2013bayesian}. The PG(1,$c$) random variable is obtained by exponential tiling of the density of a PG(1,0) random variable. Letting $p_W(w|c)$ denote the density of a PG(1,$c$) random variable,
\begin{equation}
	p_W(w|c) = \text{cosh}\left(\frac{c}{2}\right)\text{e}^{-\frac{c^2w}{2}}p_W(w).
\end{equation}

\noindent \textbf{PG augmentation preserves the Bernoulli likelihood}: Following the treatment of~\citep{choi2013polya}, conditioned on $X = x$, let $W$ be a PG(1,$|x|$) random variable. Further suppose that, conditioned on $X = x$, $\Delta N$ and $W$ are independent. Then
\begin{equation}
	p(\Delta N,w|x) = p(\Delta N|x)p_W(w|x).
\end{equation}
\noindent Integrating out $W$, we see that the augmentation scheme does not alter $p(\Delta N|x)$. One may then ask, what is the utility of the augmentation scheme? The answer lies in the following identity, discussed in detail in~\citep{choi2013polya}, and which is the key ideal behind PG augmentation
\begin{equation}
	p(\Delta N|x)p_W(w|x) = \frac{(\text{e}^x)^{\Delta N}}{1+\text{e}^x}\cdot\text{PG}(1,|x|)\propto \text{e}^{-\frac{1}{2}\frac{(\tilde{y}-x)^2}{1/w}} \propto  \mathcal{N}\left(\tilde{y};x,\frac{1}{w}\right),
	\label{eq:pggsm}
\end{equation}
\noindent where $\tilde{y} = \frac{y-\frac{1}{2}}{w}$, and $\propto$ indicates that we are dropping terms independent of $x$. Equation~\ref{eq:pggsm} states that, given $X = x$ and a logit model, a Bernoulli random variable is, up to a constant independent of $x$, a scale mixture of Gaussian~\citep{andrews1974scale}, i.e. a Gaussian random variable with random variance $1/w$, where $W = w$ follows a PG distribution~\citep{polson2013bayesian,choi2013polya}. If we assume $X \sim p_X(x)$, then~\citep{choi2013polya}
\begin{equation}
	p(\Delta N,x,w) = p(\Delta N|x,w)p_W(w|x)p_X(x) \propto \mathcal{N}\left(\tilde{y};x,\frac{1}{w}\right) p_X(x).
	\label{eq:keyid}
\end{equation}
\noindent\textbf{Implications of augmentation on $p(x|\Delta N,w)$ and $p(w|\Delta N,x)$}: 
\begin{equation}
	p(x|\Delta N,w) = \frac{p(\Delta N,x,w)}{p(\Delta N,w)} \propto p(\Delta N|x,w)p_W(w|x)p_X(x)  \propto \mathcal{N}\left(\tilde{y};x,\frac{1}{w}\right) p_X(x),
	\label{eq:xfullcond}
\end{equation}
\noindent where we make use of Equation~\ref{eq:keyid}. If $X$ is Gaussian, then $p(x|\Delta N,w)$ is Gaussian and available in closed-form! (Appendix).
\\
\begin{equation}
	p_W(w|\Delta N,x) = \frac{p(\Delta N,x,w)}{p(\Delta N,x)} = \frac{p(\Delta N|x)p_W(w|x)p_X(x)}{\int_w p(\Delta N|x)p_W(w|x)p_X(x)} = p_W(w|x),
	\label{eq:wfullcond}
\end{equation}
i.e. \noindent $p(w|\Delta N,x) = p_W(w|x) = \text{PG}(1,|x|)$. Together,  Equations~\ref{eq:xfullcond} and~\ref{eq:wfullcond} form the basis of a uniformly ergodic~\citep{choi2013polya} Gibbs sampler to obtain sample from $p(x,w|\Delta N)$.}

\subsection{Block Gibbs sampler for PG-augmented SMuRF model}

Consider the following version of the SMuRF model with PG augmentation:
\begin{equation}\left\{
\begin{array}{ll}
      {{x_k = \rho_x x_{k-1} + \alpha_x u_{x,k}+\epsilon_k, \epsilon_k \sim N(0,\sigma^2_{\epsilon})}} &  \\
      {{z_r = \rho_z z_{r-1} + \alpha_z u_{z,k} +\delta_r, \delta_r \sim N(0,\sigma^2_{\delta})}} &  \\
      \lambda_{k,r}\Delta = \frac{e^{x_k+z_r}}{1+e^{x_k+z_r}} & \\
     \Delta N_{k,r} | x_k,z_r \sim \text{Bernoulli}(\lambda_{k,r}\Delta) & \\
      w_{k,r} | x_k, z_r \sim \text{PG}(1,|x_r + z_r|), k = 1,\cdots,K; r=1,\cdots, R.
\end{array}
\right.
\label{eqn:augmented_smurf}
\end{equation}
\noindent We can apply the basic results from the previous subsection to derive the following result (proof in Appendix):
\begin{thm}
	Suppose $\w$, $\x$, $\z$ and $\mathbf{\Delta N}$ come from the PG-augmented SMuRF model (equation ), then $p(\Delta N,\x|\w,\z;\theta)$ is equivalent in distribution to the following linear-Gaussian state-space model
\begin{equation} \left\{\begin{array}{ll}
	{{x_k = \rho_x x_{k-1} + \alpha_xu_{x,k} +\epsilon_k, \epsilon_k \sim \mathcal{N}(0,\sigma^2_{\epsilon})}} & \\
        \Delta \tilde{N}_{k,r} = x_k + z_r + \tilde{v}_{k,r}, \tilde{v}_{k,r} \sim \mathcal{N}(0, {w_{k,r}}^{-1}), \text{ i.i.d. }, r = 1,\cdots,R & \\
      \Delta \tilde{N}_{k,r} = \frac{\Delta N_{k,r} - \frac{1}{2}}{w_{k,r}}.&
\end{array}
\right .\
\label{eqn:eqv_1d_model}
\end{equation}
\label{theorem2}
\end{thm}
\noindent Following the discussion from the previous subsection, it is not hard to see that such a result would hold. The proof of this result is in the appendix, as well as the derivation of an elegant forward-filtering backward-sampling algorithm~\citep{fruhwirth1994data} for drawing samples from $p(\Delta N,\x|\w,\z;\theta)$. By symmetry, it is not hard to see that a similar result holds for $p(\Delta N,\z|\w,\x;\theta)$.

\paragraph{\underline{Block Gibbs sampling from PG-augmented SMuRF model:}}

The E-step of the Monte-Carlo EM algorithm consists in sampling from $p(\x,\z|\mathbf{\Delta N};\theta^{(\ell)})$ by drawing from $p(\x,\z,\w|\Delta N;\theta^{(\ell)})$ using a block Gibbs sampler that uses the following full-conditionals
\begin{itemize}
	\item[$\bullet$] $p(\x|\mathbf{\Delta N},\w,\z;\theta^{(\ell)})$, which according to the theorem above is equivalent to the posterior distribution of the state sequence in a linear-Gaussian state-space model.
	\item[$\bullet$] $p(\z|\mathbf{\Delta N},\w,\x;\theta^{(\ell)})$, which obeys properties similar to the previous full-conditional (by symmetry).
	\item[$\bullet$] $p(w_{k,r}|\x,\z) = p(w_{k,r}|x_k,z_r) = \text{PG}(1,|x_k + z_r|), k = 1,\cdots, K, r = 1,\cdots, R$~\citep{polson2013bayesian}.
\end{itemize}

\noindent In the Appendix, we detail how we initialize the algorithm and monitor convergence.

{{In practice, we found that estimating $\rho_x$ and $\rho_z$ is difficult. We hypothesize that including those parameters yields an unwieldy likelihood function. In the results we report, we assume $\rho_x = \rho_z = 1$, $\alpha_x = \alpha_z = 0$ and focus on estimating a simple model with two parameters $\sigma^2_\epsilon$ and $\sigma^2_\delta$. The assumption $\rho_x = \rho_z = 1$ gives the random walk priors more freedom, thus allowing us to be capture the variability of the within and cross-trial processes. We have run simulations, not reported here, that show that the joint estimation of $\sigma^2_\delta$, $\sigma^2_\epsilon$, $\alpha_x$ and $\alpha_z$ is stable and that our EM algorithm converges. This demonstrates the ability of the SMuRF model {{(Equation~(\ref{eqn:augmented_smurf}))}} to incorporate exogenous input stimuli.}}

\noindent {{Assuming $\rho_x = \rho_z = 1$, in the M-step, the update equations for the parameters $\sigma^2_\epsilon$, $\sigma^2_\delta$, $\alpha_x$ and $\alpha_z$}} follow standard formulas~\citep{smith2003estimating}
\begin{eqnarray}
	{{\alpha^{(\ell+1)}_{x}}} & = & {{\frac{\sum_{k=1}^K\left(\mathbb{E}_{\mathbf{x}}\left[x_k - x_{k-1}|\mathbf{\Delta N},\theta^{(\ell)}\right]\right)u_{x,k}}{\sum_{k=1}^K u_{x,k}^2}}},\\
	{{\alpha^{(\ell+1)}_{z}}} & = & {{\frac{\sum_{r=1}^R\left(\mathbb{E}_{\mathbf{z}}\left[z_r - z_{r-1}|\mathbf{\Delta N},\theta^{(\ell)}\right]\right)u_{z,r}}{\sum_{r=1}^R u_{z,r}^2}}},\\
	\sigma^{2(\ell+1)}_{\epsilon} & = & \mathbb{E}_{\mathbf{x}}\left[\frac{1}{K}\sum_{k=1}^K{{(x_k-x_{k-1}-\alpha_x^{(\ell+1)} u_{z,k})^2}}|\mathbf{\Delta N},\theta^{(\ell)}\right],\\
	\sigma^{2(\ell+1)}_{\delta} & = & \mathbb{E}_{\mathbf{z}}\left[\frac{1}{R}\sum_{r=1}^R{{(z_r-z_{r-1}-\alpha_z^{(\ell+1)} u_{z,k})^2}}|\mathbf{\Delta N},\theta^{(\ell)}\right],
	\
\end{eqnarray}
\noindent where we set $x_0 = z_0 = 0$, and we approximate the expectations with respect to $p(\x|\Delta N,\theta^{(\ell)})$ and $p(\z|\Delta N,\theta^{(\ell)})$ using Gibbs samples from the E-step.


\subsection{Assessment of within-trial and cross-trial spiking dynamics}

Bayesian estimation of the SMuRF model {{(Equation~(\ref{eqn:augmented_smurf}))}} enables us to infer detailed changes in neural dynamics, in particular to extract the within-trial and cross-trial components of the neural spiking dynamics that accompany the learning of a contingency by a neuron. This is because, following estimation, inference in the SMuRF model yields the \emph{joint} posterior distribution of the instantaneous spiking rate of a neuron as a function of trials, and time within a trial, conditioned on the observed data. We can use this posterior distribution, in turn, to assess \emph{instantaneous} changes in neural spiking dynamics, and \emph{without the need to correct for multiple comparisons} as with non-parametric methods.
\\
\\
\noindent In what follows, we let $p(\x,\z|\mathbf{\Delta} N;\hat{\theta}_{ML})$ denote the posterior distribution of $\x$ and $\z$, given the raster data $\mathbf{\Delta N}$ and the maximum likelihood estimate $\hat{\theta}_{ML}$ of $\theta$. In what follows, it is understood that we use Gibbs samples $(\x_i,\z_i)_{i=1}^n$ from $p(\x,\z|\mathbf{\Delta} N;\hat{\theta}_{ML})$ to obtain an empirical estimate of the distribution.

\noindent \underline{\textbf{Posterior distribution of the joint CIF over time and trials}}:  We can use these posterior samples to approximate the posterior distribution, at $\hat{\theta}_{ML}$, of any quantities of interest. Indeed, it is well known from basic probability that if $(\x_i,\z_i)$ is a sample from $p(\x,\z|\mathbf{\Delta} N;\hat{\theta}_{ML})$, then $f(\x_i,\z_i)$ is a sample from $p(f(\x,\z)|\mathbf{\Delta} N;\hat{\theta}_{ML})$. In particular, if the instantaneous spiking rate of a neuron a time $k$ and trial $r$ $\lambda_{k,r}\Delta = \frac{e^{x_k+z_r}}{1+e^{x_k+z_r}}$, we can use the Gibbs samples to approximate the joint posterior distribution
of $\{\lambda_{k,r}\Delta\}_{k=1,r=1}^{K,R}$ given $\mathbf{\Delta N}$ and $\hat{\theta}_{ML}$.
\\

\noindent Let $\{\lambda^\text{p}_{k,r}\Delta\}_{k=1,r=1}^{K,R}$ be the random variable that represents the \emph{a posteriori} instantaneous spiking rate of the neuron at time trial $r$ and time $k$ within that trial. The superscript `p' highlights the conditioning on the data $\mathbf{\Delta N}$ and $\hat{\theta}_{ML}$, and the fact that this quantity is a function of $(\x,\z)$ distributed according to $p(\x,\z|\mathbf{\Delta} N;\hat{\theta}_{ML})$.


\noindent \underline{\textbf{Within-trial effect}}: We define the within-trial effect as the \emph{a posteriori} instantaneous spiking rate at time $k$, average over all trials
\begin{equation}
	e^{\text{WT}}_k = \frac{1}{R} \sum_{r=1}^R \lambda^\text{p}_{k,r} (x_k,z_r), k = 1,\cdots,K.
\end{equation}
\noindent It is important to note that the averaging is performed \emph{after} characterization of the joint CIF as a function of time and trials, which is not the same as first aggregating the data across trials and applying one of the one-dimensional methods for analyzing neural data~\citep{smith2003estimating,Zammit:2012,Yuan:2012}. In practice, every Gibbs sample pair $(\x_i,\z_i), i=1,\cdots,n$ leads to a scalar quantity
\begin{equation}
	\hat{e}^{\text{WT}}_{i,k} = \frac{1}{R} \sum_{r=1}^R \lambda_{k,r} (\x_{i,k},\z_{i,r}), k = 1,\cdots,K.
\end{equation}

\noindent Performing this computation over all Gibbs samples and times $k=1,\cdots,K$ leads to a joint empirical distribution for the within-trial effect $\{e^{WT}_k\}_{k=1}^K$.

\noindent \underline{\textbf{Cross-trial effect}}: We define the cross-trial effect as the \emph{a posteriori} excess instantaneous spiking at trial $r$ and time $k$ (above the within-trial effect effect $e^{WT}_k$) averaged across all times $k$
\begin{equation}
	e^{\text{CT}}_r = \frac{1}{K} \sum_{k=1}^K \frac{\lambda^p_{k,r} (x_k,z_r)}{e^{WT}_k}, r = 1,\cdots,R.
	\label{eq:cte}
\end{equation}

\noindent In practice, every Gibbs sample pair $(\x_i,\z_i), i=1,\cdots,n$ leads to a scalar quantity
\begin{equation}
	\hat{e}^{\text{CT}}_{i,r} = \frac{1}{K} \sum_{k=1}^K \frac{\lambda_{k,r} (\x_{i,k},\z_{i,r})}{\hat{e}^{WT}_{i,k}}, r = 1,\cdots,R.
\end{equation}

\noindent Performing this computation over all Gibbs samples and $R_c$ trials of interest, $r=R-R_c+1,\cdots,R$, leads to a joint empirical distribution for the cross-trial effect $\{e^{\text{CT}}_r\}_{r=R-R_c}^R$.
\\
\noindent \underline{\textbf{Remark 4}}: The following paragraph explains the meaning of $R_c$ in the context of an associative learning experiment.

\subsection{Assessment of neural spiking dynamics across time and trials}

\begin{figure}[h!]
        \begin{center}
        \includegraphics[scale=1]{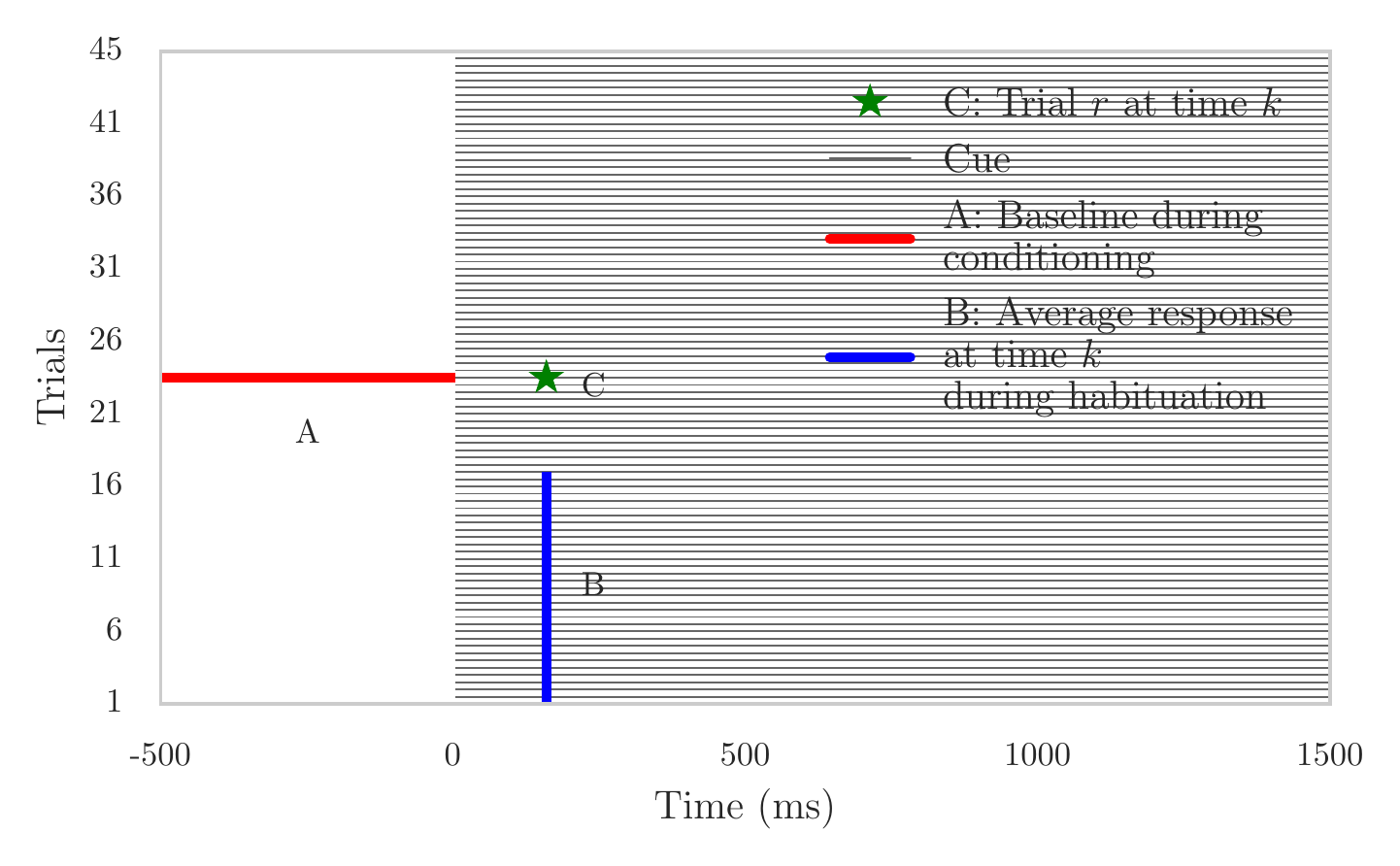}
        \end{center}
    \vspace*{-3mm}
    \caption{{Regions defined to quantify changes in neural spiking dynamics in an associative learning experiment. The SMuRF model lets us approximate the \emph{a posteriori} probability that the spiking rate at a given point C during one of the conditioning trials (trials $16$ through $45$ in this example) is bigger than the baseline spiking rate at that trial (Region A) and the average spiking rate at the same time during the habituation period (Region B). This yields a probabilistic description of the the intricate dynamics of neural spiking that accompany the learning of the contingency in the experiment by a neuron.}}
    \label{fig:region}
\end{figure}

\noindent  Consider an associative learning (conditioning) experiment characterized by the pairing of a conditioned stimulus (e.g. auditory) to an aversive stimulus (e.g. a shock). Let $R_c$ be the number of conditioning trials and $K_h$ the length of the habituation period. Gibbs samples from the SMuRF model {{(Equation~(\ref{eqn:augmented_smurf}))}} paramaterized by $\hat{\theta}_{ML}$ let us approximate the \emph{a posteriori} probability that the spiking rate at a given point (Point C in Figure \ref{fig:region}) during one of the conditioning trials (trials $16$ through $45$ in this example) is bigger than the baseline spiking rate at that trial (Region A in Figure \ref{fig:region}) and the average spiking rate at the same time during the habituation period (Region B in Figure \ref{fig:region}). This yields a probabilistic description of the the intricate dynamics of neural spiking that accompany the learning of the contingency by a neuron. Let
\begin{eqnarray}
\text{Event U}  &=  \left\{\lambda^p_{k,r} (x_k,z_r) > \overbrace{\frac{1}{R_c} \sum_{m=1}^{R_c} \lambda^\text{p}_{k,m} (x_k,z_m)}^{\text{Average rate in Region A}}\right\} \\
\text{Event V}  &= \left\{\lambda^\text{p}_{k,r} (x_k,z_r) > \overbrace{\frac{1}{K_h} \sum_{s=1}^{K_h} \lambda^\text{p}_{s,r} (x_s,z_r)}^{\text{Average rate in Region B}}\right\}
\end{eqnarray}
For a given pair $(k,r)$ s.t. $k \geq K_h, r \geq R_c$, this probability is
\begin{eqnarray}
	 & & \mathbb{P}\left[\text{Event U} \cap \text{Event V}\right] \\ & \approx & \frac{1}{n} \sum\limits_{i=1}^n \mathbb{I}_{\{\lambda^\text{p}_{k,r} (\x_{i,k},\z_{i,r}) > \frac{1}{R_c} \sum\limits_{m=1}^{R_c} \lambda^\text{p}_{k,m} (\x_{i,k},\z_{i,m}) \cap \lambda^\text{p}_{k,r} (\x_{i,k},\z_{i,r}) > \frac{1}{K_h} \sum\limits_{s=1}^{K_h} \lambda^\text{p}_{s,r} (\x_{i,s},\z_{i,r})\}},
	 \label{eq:probev}
\end{eqnarray}
\noindent where the second line approximates the probability of the event of interest using its frequency of occurrence in the $n$ posterior samples. As we demonstrate in the following section, we thus obtain an detailed characterization of the dynamics of neural spiking that accompany learning.


In the following section, we use simulated and real data examples to demonstrate the utility of the SMuRF model {{(Equation~(\ref{eqn:augmented_smurf}))}} for the characterization of detailed neural spiking dynamics.

\section{Applications}
\label{sec:res}

\subsection{Simulation studies}

\begin{figure}[h!]
        \begin{center}
        \includegraphics[scale=1]{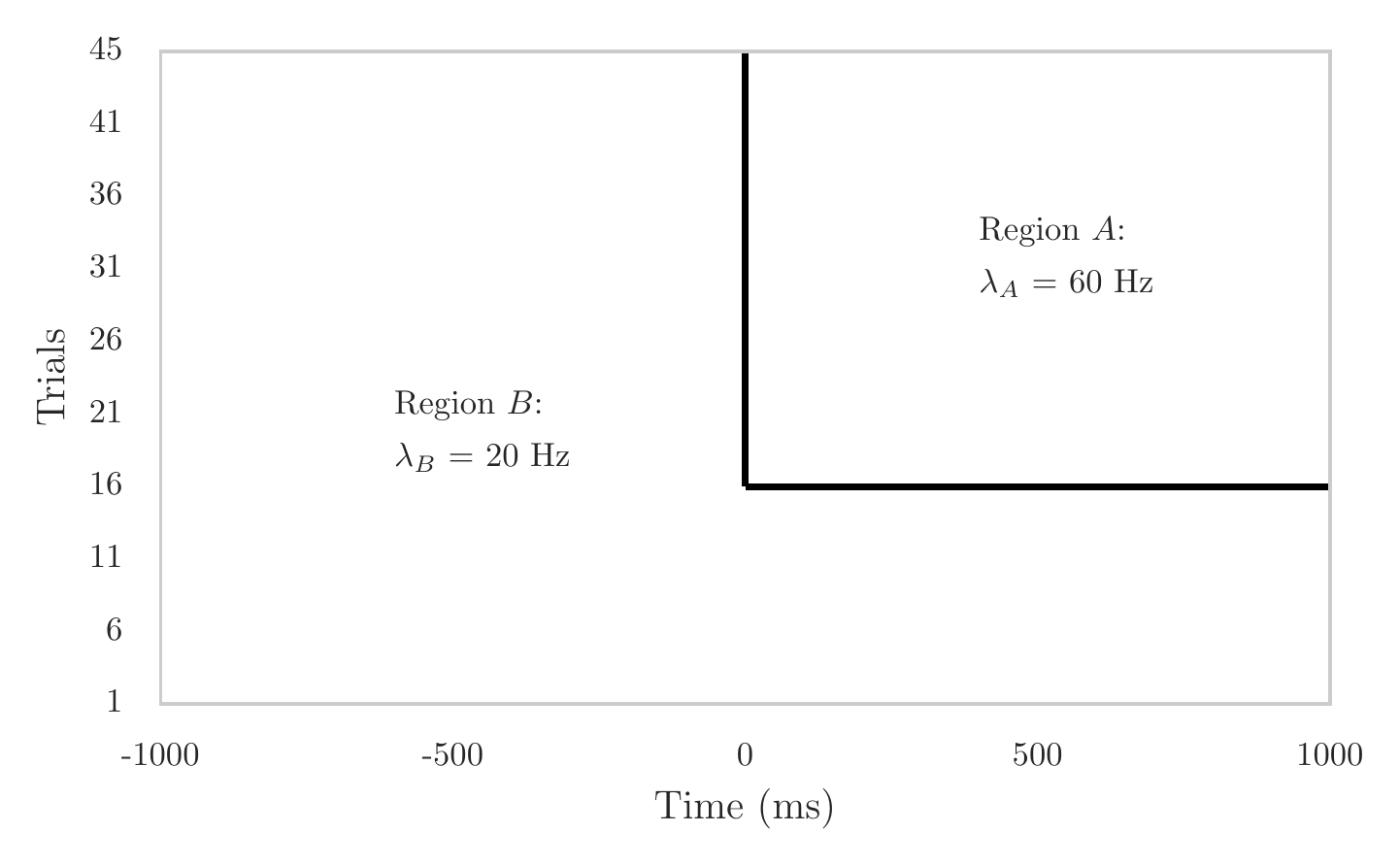}
        \end{center}
    \vspace*{-3mm}
    \caption{{Set up used to simulate neural spike raster data from a neuron that exhibits a conditioned response to the conditioned stimulus in an associative learning experiment. The experiment consists of $45$ trials, each of which lasts $2$ s. The conditioned stimulus becomes active $1$ s into a trial, while the aversive stimulus becomes active from the $16^{\text{th}}$ trial onwards. An idealized neuron that exhibits a conditioned response would exhibit two distinct regions of activity. Region A consists of all trials preceding trial $16$, along with the period from all trials \emph{before} the conditioned stimulus is presented. We assume that the rate of spiking of the neuron is $\lambda_A = 60$ Hz. Region B consists of the period from trials following trial $15$ \emph{after} the conditioned stimulus is presented. The rate of spiking of the neuron in this region is $\lambda_B = 20$ Hz.}}
    \label{fig:simulated_region}
\end{figure}

\begin{figure}[h!]
        \begin{center}
        \includegraphics[scale=1]{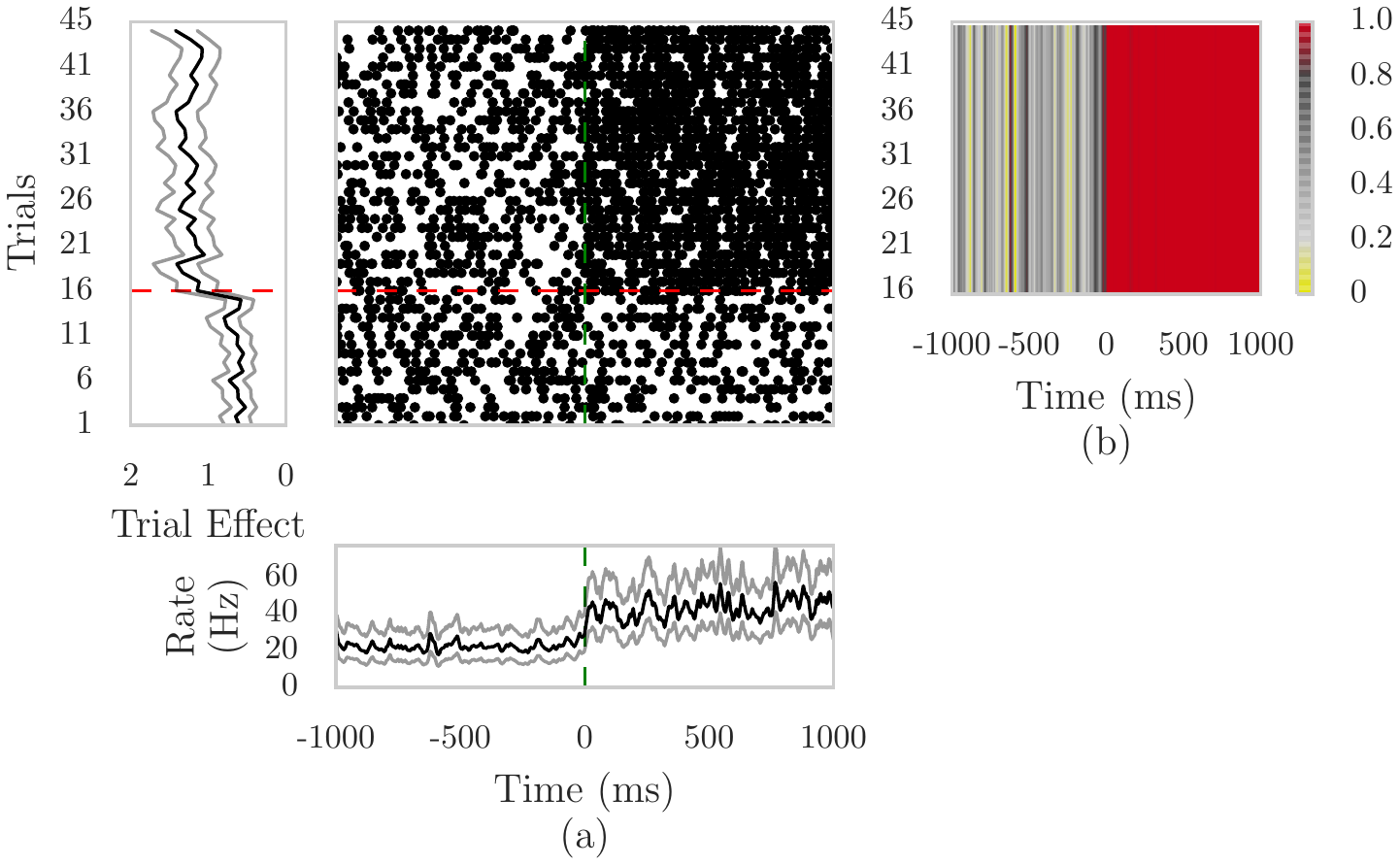}
        \end{center}
    \vspace*{-3mm}
    \caption{{(a) Simulated neural spike raster, along with estimated within and cross trial effects from the SMuRF model. The horizontal red line indicates the beginning of conditioning. The vertical green line indicates the onset of the conditioned stimulus. The left panel of the figure shows the cross-trial effect which, following conditioning, increases above its average initial value of $\approx 1$ to $\approx 2$. (b) Empirical probability that spiking rate at a given trial and time is bigger than the average rate at the same trial and the average rate during habituation at the same time. With probability close to $1$, the spiking rate at a given time/trial pair--following the conditioned stimulus and during conditioning (Figure~\ref{fig:region} C)--is bigger than the average rate at the same trial (Figure~\ref{fig:region} A) and the average rate at the same time (Figure~\ref{fig:region} B)}}
    \label{fig:simulated_data}
\end{figure}

\noindent We simulated neural spike raster data from a neuron that exhibits a conditioned response to the conditioned stimulus (Figure~\ref{fig:simulated_data}) in an associative learning experiment. The experiment consists of $45$ trials, each of which lasts $2$ s. The conditioned stimulus becomes active $1$ s into a trial, while the aversive stimulus becomes active after trial $15$. We obtain the simulated data by dividing the raster into two pre-defined regions as shown in Figure~\ref{fig:simulated_region}. Region A consists of all trials before trial $16$, along with the period from all trials \emph{before} the conditioned stimulus is presented. We assume that the rate of spiking of the neuron is $\lambda_A = 60$ Hz. Region B consists of the period from trials following trial $15$ \emph{after} the conditioned stimulus is presented. The rate of spiking of the neuron in this region is $\lambda_B = 20$ Hz. 

We applied the SMuRF model {{(Equation~(\ref{eqn:augmented_smurf}))}} to the analysis of this simulated neural spike raster. Figure~\ref{fig:simulated_data}(a) shows that, during conditioning, learning is accompanied by a doubling of the spiking rate above the within-trial spiking rate of the neuron. Indeed, the left hand panel of the figure shows the cross-trial effect which, following conditioning, increases above its average initial value of $\approx 1$ to $\approx 2$. Figure~\ref{fig:simulated_data}(b) provides a more detailed characterization of the neural spiking dynamics. With probability close to $1$, the spiking rate at a given time/trial pair--following the conditioned stimulus and during conditioning (Figure~\ref{fig:region} C)--is bigger than the average rate at the same trial (Figure~\ref{fig:region} A) and the average rate at the same time (Figure~\ref{fig:region} B). We conclude that, with high probability, the simulated neuron exhibits a conditioned response to the conditioned stimulus.

\subsection{Neural dynamics during associative learning of fear}

\paragraph{Basic experimental paradigm:} The ability to learn through observation is a powerful means of learning about aversive stimuli without direct experience of said stimuli. We use the SMuRF model {{(Equation~(\ref{eqn:augmented_smurf}))}} to analyze data from a fear conditioning paradigm designed to elucidate the nature of the circuits that facilitate the associative learning of fear. The experimental paradigm is described in detail in~\citep{allsop2017observational}. Briefly, an observer mouse observes a demonstrator receive conditioned stimulus-shock pairings through a perforated transparent divider. The experiment consists of $45$ to $50$ trials, divided into two phases. During the first $15$ trials of the experiment, termed the habituation period, both the observer and the demonstrator simply hear an auditory conditioned stimulus. {{From the $16^{\text{th}}$ trial onwards}}, the auditory conditioned stimulus is followed by the delivery of a shock to the demonstrator. The data are recorded from the pre-frontal cortex (PFC) of the observer mouse.

\begin{figure}[H]
        \begin{center}
        \includegraphics[scale=1]{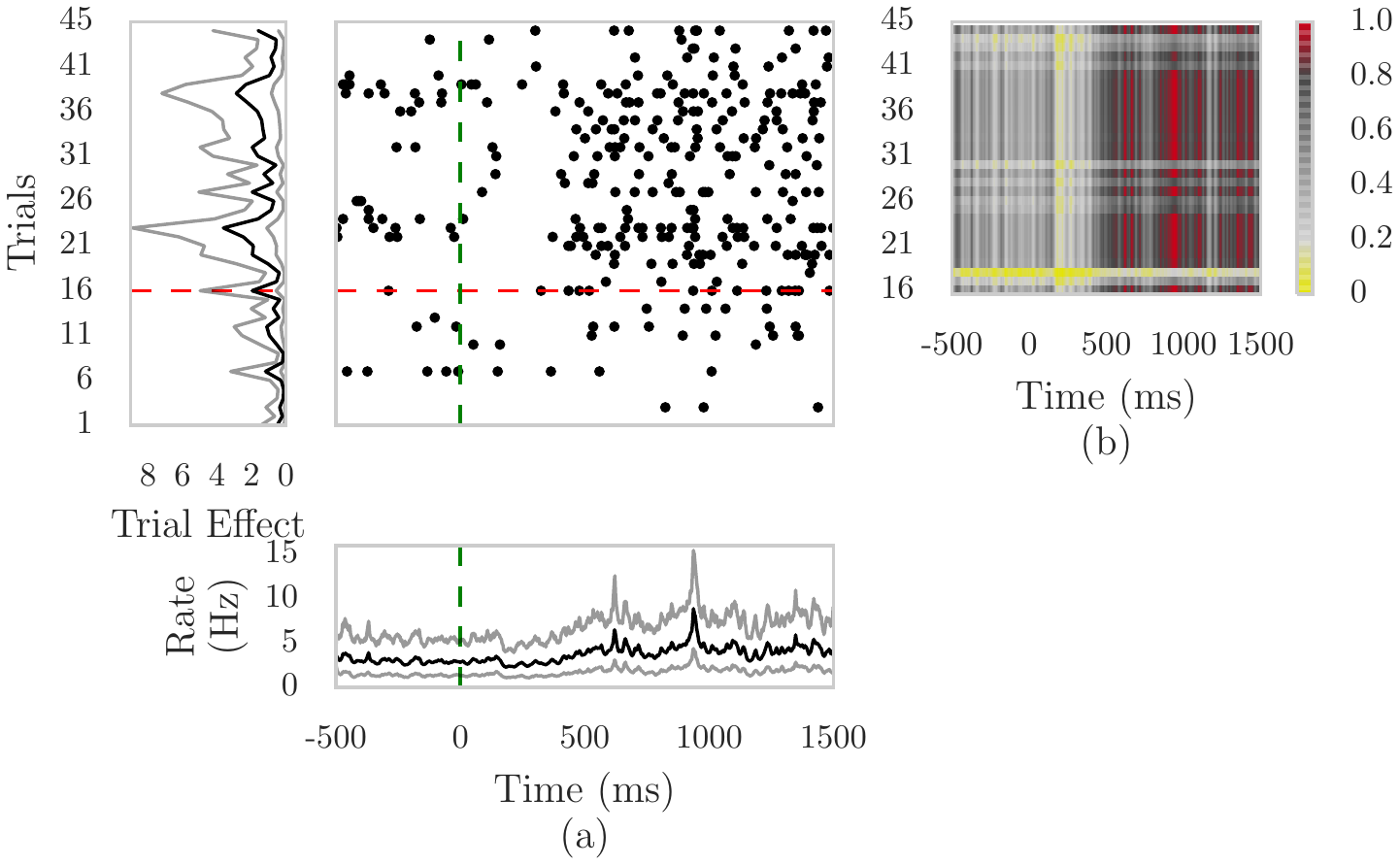}

        \end{center}

    \vspace*{-3mm}

    \caption{{(a) Raster, along with SMuRF within and cross-trial components, from a cortical neuron that exhibits a conditioned response to the conditioned stimulus. The horizontal red line indicates the beginning of conditioning. The vertical green line indicates conditioned stimulus onset. The bottom panel suggests that this neuron exhibits a delayed response to the conditioned stimulus, beginning at $\approx 400$ ms following conditioned stimulus presentation. Accounting for this increase in within-trial spiking rate due to the conditioned stimulus, the left panel shows a multiplicative increase in spiking rate due to conditioning from an average initial value of $\approx < 1$ (indicative of suppression, as can be seen through the sparseness of the raster during trials $1$ through $5$) to a peak average value of $\approx 4$ at trial $23$. This increase, however, does not persist as in the case of the simulated data (Figure~\ref{fig:simulated_data}).}}

    \label{fig:new_raster_2192}
\end{figure}

\paragraph{Results:}  Figure~\ref{fig:new_raster_2192}(a) shows the within and cross-trial effects estimated using the SMuRF model applied to a cortical neuron from the experiment described above. The estimated within-trial (bottom) and cross-trial (left) components indicate significant changes respectively in response to the conditioned stimulus and to conditioning. By definition (Equation \ref{eq:cte}), the cross-trial effects takes into account the increase in spiking rate due to the presentation of the conditioned stimulus. The bottom panel suggests that this neuron exhibits a delayed response to the conditioned stimulus, beginning at $\approx 400$ ms following conditioned stimulus presentation. Accounting for this increase in within-trial spiking rate due to the conditioned stimulus, the left panel shows a multiplicative increase in spiking rate due to conditioning from an average initial value of $\approx < 1$ (indicative of suppression, as can be seen through the sparseness of the raster during trials $1$ through $5$) to a peak average value of $\approx 4$ at trial $23$. This increase, however, does not persist as in the case of the simulated data (Figure~\ref{fig:simulated_data}), suggesting that conditioning is accompanied by intricate dynamics in neural modulation.
\addtocounter{figure}{-1}
\begin{figure} [h!]
    \captionof{figure}{{(b) Empirical probability that spiking rate at a given trial in and time is bigger than the average rate at the same trial and the average rate during habituation at the same time (refer to Figure~\ref{fig:region}). This panel indicates that this neuron exhibit a delayed conditioning to the conditioned stimulus (beginning $\approx 400$ ms following conditioned stimulus presentation) and that the extent of the condition is highest first between trials $18$ and $24$ and then between trials $31$ and $41$. Panel (a), and panel (b) in particular, suggest that conditioning is accompanied by intricate dynamics in neural modulation.}}
\end{figure}

Figure~\ref{fig:new_raster_2192} (b) provides a more detailed characterization of the neural spiking dynamics of this neuron.  The figure shows the evolution, as a function of time and trials, of the probability that the spiking rate at a given time/trial pair (Figure~\ref{fig:region} C) is bigger than the average rate at the same trial (Figure~\ref{fig:region} A) and the average rate at the same time (Figure~\ref{fig:region} B). The figure indicates that this neuron exhibit a delayed conditioning to the conditioned stimulus (beginning $\approx 400$ ms following conditioned stimulus presentation) and that the extent of the condition is highest first between trials $18$ and $24$ and then between trials $31$ and $41$.

%
%
%
%
%
%

\begin{figure}[H]
        \begin{center}
        \includegraphics[scale=1]{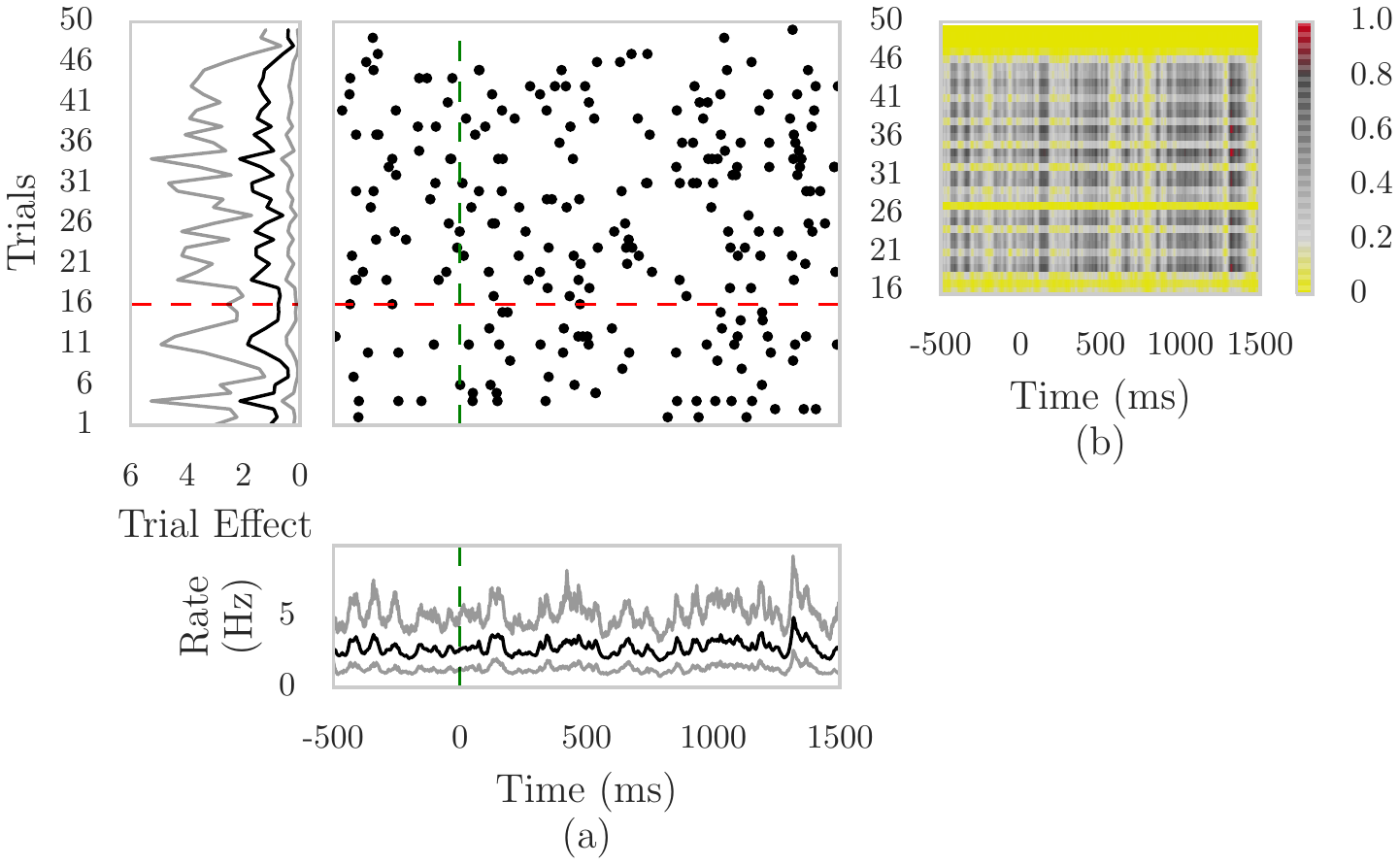}

        \end{center}

    \vspace*{-3mm}

    \caption{{(a) Raster, along with SMuRF within and cross-trial components, from a cortical neuron that does not exhibit a conditioned response to the conditioned stimulus. The horizontal red line indicates the beginning of conditioning. The vertical green line indicates the onset of the conditioned stimulus. This neuron does not exhibit a conditioned response to the conditioned stimulus. The bottom of the panel indicates no significant increase in the within-trial spiking rate in response to the conditioned stimulus, while the left panel shows that the cross-trial effect remains constant throughout the experiment an average value of $\approx 1$. This indicates that conditioning does not result in a significant increase in spiking rate.}}

    \label{fig:new_raster_2119}

\end{figure}

Figure~\ref{fig:new_raster_2119} shows an application of the SMuRF model {{(Equation~(\ref{eqn:augmented_smurf}))}} to a cortical neuron that does not exhibit a conditioned response to the conditioned stimulus. The bottom of panel (a) indicates no significant increase in the within-trial spiking rate in response to the conditioned stimulus, while the left panel shows that the cross-trial effect remains constant throughout the experiment an average value of $\approx 1$. This indicates that conditioning does not result in a significant increase in spiking rate. Panel (b) corroborates these findings: for all points C following the conditioned stimulus and during conditioning, there is a small probability that the instantaneous spiking rate is significantly different from the average spiking rates in Regions A and B.

\addtocounter{figure}{-1}
\begin{figure} [h!]
    \captionof{figure}{{(b) Empirical probability that spiking rate at a given trial and time is bigger than the average rate at the same trial and the average rate during habituation at the same time This panel corroborates the observations from panel (a): for all points C following the conditioned stimulus and during conditioning, there is a small probability that the instantaneous spiking rate is significantly different from the average spiking rates in Regions A and B.}}
\end{figure}

\noindent Figures~\ref{fig:new_raster_2193} and~\ref{fig:new_raster_3189}  show results for two additional cortical neurons that exhibit a \emph{transient} conditioned response to the conditioned stimulus.

{

\paragraph{Using SMuRF inference to determine a neuron's learning time and trial}

The power of the Bayesian approach, and the SMuRF model {{(Equation~(\ref{eqn:augmented_smurf}))}} in particular, lies in the fact that it lets us approximate the \emph{a posteriori} probability that the spiking rate at a given point (point C in Figure \ref{fig:region}) during one of the conditioning trials (trials $16$ through $45$ in this example) is bigger than the baseline spiking rate at that trial (Region A in Figure \ref{fig:region}) and the average spiking rate at the same time during the habituation period (Region B in Figure \ref{fig:region}) (Equation~\ref{eq:probev}). This yields an instantaneous probabilistic quantification of the extent of learning for any given time and trial pair.

Panel (b) of Figures~\ref{fig:new_raster_2192},~\ref{fig:new_raster_2119},~\ref{fig:new_raster_2193} and~\ref{fig:new_raster_3189} provide a detailed characterizations of the dynamics of learning and its extent for all times following the onset of the conditioned stimulus, \emph{all} conditioning trials.

Here, we provide some guidance for practitioners to summarize the results of our inference to a single learning time/trial pair. We would like to stress, however, that the power of our methods lies in the detail provided by panel (b) of Figures~\ref{fig:new_raster_2192},~\ref{fig:new_raster_2119},~\ref{fig:new_raster_2193} and~\ref{fig:new_raster_3189}. Since the SMuRF model enables us to compute a empirical probability that the spiking rate at a given time/trial pair (Figure~\ref{fig:region}, Point C) is bigger than the average rate at the same trial (Figure~\ref{fig:region}, Region A) and the average rate at the same time (Figure~\ref{fig:region}, Region B), we can identify learning time and learning trial for each neuron by finding the first time after cue and after conditioning that this probability exceeds a certain threshold. Table ~\ref{learning_time_trial_table} reports the learning time and trial computed using a threshold of of 95\%. Note that the learning time is computed with respect to the onset of the conditioned stimulus (time $=0$ ms).
\begin{table}[h]
\centering
\begin{tabular}{ |p{2cm}|p{2cm}|p{2cm}|p{2cm}|p{2cm}|}
 \hline
  & Neuron in Figure~\ref{fig:new_raster_2192} & Neuron in Figure~\ref{fig:new_raster_2119}& Neuron in Figure~\ref{fig:new_raster_2193} & Neuron in Figure~\ref{fig:new_raster_3189} \\ \hline
 Learning time (ms) & 617 & 1316 & 202 & 20 \\ \hline
 Learning trial          & 16 & 34 & 18 & 18 \\
 \hline
\end{tabular}
 \caption{{Learning trial and time computed for the cortical neurons analyzed. The learning times and trials reported are consistent with the detailed inference provided by the respective Figures for these neurons. The cortical unit from Figure~\ref{fig:new_raster_2192}, for instance, shows a delayed response, significant $617$ ms after conditioned stimulus onset and at trial $16$. The cortical unit from Figure~\ref{fig:new_raster_2119} only exhibits a significant change in neural spiking $1316$ ms following the conditioned stimulus and at trial $34$. This is consistent with our previous observation from Figure~\ref{fig:new_raster_2119} that this neuron does not exhibit a conditioned response to the stimulus.}}
 \label{learning_time_trial_table}
\end{table}

The learning times and trials reported in Table~\ref{learning_time_trial_table} are consistent with the detailed inference provided by the respective Figures for these neurons. Indeed, the cortical unit from Figure~\ref{fig:new_raster_2192} shows a delayed response, significant $617$ ms after conditioned stimulus onset and at trial $16$. The cortical unit from Figure~\ref{fig:new_raster_2119} only exhibits a significant change in neural spiking $1316$ ms following the conditioned stimulus and at trial $34$. This is consistent with our previous observation from Figure~\ref{fig:new_raster_2119} that this neuron does not exhibit a conditioned response to the stimulus.

In the Appendix, we perform a simulation that demonstrates the ability of SMuRF inference to identify learning time and trial when learning of a contingency is accompanied by \emph{sustained} changes in neural spiking following conditioned stimulus onset and during conditioning. We also demonstrate through simulation that the SMuRF model is robust to the presence of error trials.

\subsection{Application of SMuRF model to a non-separable example}

We demonstrate the limitations of the separability assumption in the SMuRF model {{(Equation~(\ref{eqn:augmented_smurf}))}} by applying it to the neural spike raster data from~\citep{Czanner:08}. We briefly describe the experiment here and refer the reader to~\citep{wirth2003single} for a more detailed description. Panel (a) of Figure~\ref{fig:czannerdata} shows neural spiking activity from a hippocampal neuron recorded during an experiment designed for a location-scene association learning task. The same scene was shown to a Macaque monkey across 55 trials, and each trial lasted 1700 ms. The first 300 ms of every trial is fixation period, and the scene is presented to the monkey from 300 to 500 ms. A delay period takes place from 800 to 1500 ms, followed by a response period from 1500 to 1700 ms.

The data from the experiment are shown in the center of panel (a) from Figure~\ref{fig:czannerdata}. The raster suggests that the time and trial-dependent CIF of this neuron is \emph{not} separable. this Intuitively, this can be seen from the fact that the region in which there are significant changes in neural spiking does not follow the rectangular form from Figure~\ref{fig:simulated_region}. Nevertheless, the CIF could be well approximated by a separable model. We apply the SMuRF model to these data to uncover some of its limitations in non-separable settings. The bottom panel of Figure~\ref{fig:czannerdata}(a) shows the estimate of the within-trial effect from the SMuRF model, while the left panel shows the cross-trial effect. These two figures indicate that the SMuRF model is able to capture within and cross-trial dynamic changes in the spiking activity of the neuron. Figure ~\ref{fig:czannerdata}(b) shows the estimate of the \emph{a-posteriori} mean instantaneous spiking rate $\{\hat{\lambda}^\text{p}_{k,r}\Delta\}_{k=1,r=1}^{K,R}$ (in Hz) of the neuron at time trial $r$ and time $k$ within that trial. This figure shows that, while the SMuRF model is able to characterize the detailed changes in spiking dynamics, it does not fully capture the non-separable nature of the raster data.

\noindent \underline{\textbf{Remark 5}}: Unlike for the cortical neurons, this experiment does not have a conditioning period. That's why, it does not make sense to generate plots such as Figure~\ref{fig:new_raster_2192}(b).

\begin{figure}[H]
        \begin{center}
		\includegraphics[scale=1]{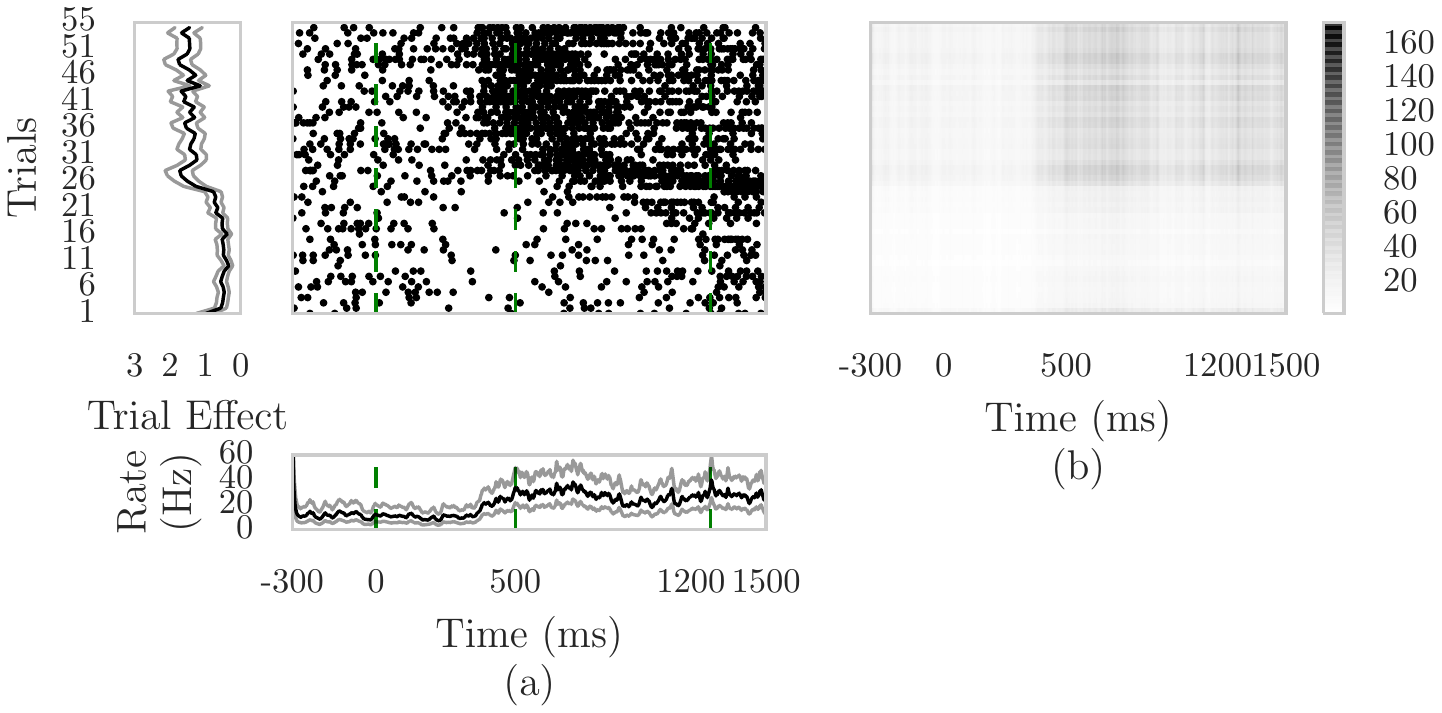}
        \end{center}
    \vspace*{-3mm}
 \caption{{Application of the SMuRF model to neural spiking activity from a hippocampal neuron recorded during an experiment designed for a a location-scene association learning task.}}
    \label{fig:czannerdata}

\end{figure}
\addtocounter{figure}{-1}
\begin{figure} [h!]
    \captionof{figure}{{The green lines represent the end of the fixation, scene presentation and delay periods respectively. (a) Neural spike raster from the hippocampal neuron. The bottom panel shows the estimate of the within-trial effect from the SMuRF model, while the left panel shows the cross-trial effect. (b) Estimate of the \emph{a-posteriori} instantaneous spiking rate $\{\hat{\lambda}^\text{p}_{k,r}\Delta\}_{k=1,r=1}^{K,R}$ (in Hz) of the hippocampal neuron at time trial $r$ and time $k$ within that trial. This figure shows that, while the SMuRF model is able to characterize the detailed changes in spiking dynamics of this neuron during the task, it fails to capture the non-separable nature of the raster data.}}
\end{figure}

}
\section{Conclusion}
\label{sec:disco}

We proposed a 2D separably-Markov random field (SMuRF) for the analysis of neural spike rasters that obviates the need to aggregate data across time or trials, as in classical one-dimensional methods~\citep{smith2003estimating,Zammit:2012,Yuan:2012}, while retaining their interpretability. The SMuRF model approximates the trial-dependent conditional intensify function (CIF) of a neuron as the product of a within-trial component, in units of Hz (spikes/s), and a unitless quantity, which we call the cross-trial effect, that represents the excess spiking rate above what can be expected from the within-trial component at that trial. One key advantage of our 2D model-based approach over non-parametric methods stems from the fact that it yields a characterization of the joint posterior (over all trials and times within a trial) distribution of the instantaneous rate of spiking of as a function of both time and trials given the data. This not only obviates the need to correct for multiple comparisons, but also enables us to compare the instantaneous
rate of any two trial time pairs at the millisecond resolution, where non-parametric methods break down because the sample size is 1.

We applied the SMuRF model to data collected from neurons in the pre-frontal cortex (PFC) in an experiment designed to characterize the neural underpinnings of the associative learning of fear in mice. We found that, as a group, the recorded cortical neurons exhibit a conditioned response to the auditory conditioned stimulus, occurring $3$ to $4$ trials into conditioning. We also found intricate and varied dynamics of the extent to which the cortical neurons exhibit a conditioned response (e.g. delays, short-term conditioning). This is likely reflective of the variability in synaptic strength, connectivity and location of the neurons in the population.

In future work, we plan to investigate non-separable random field models of neural spike rasters, such as Markov random fields~\citep{besag1974spatial} (MRFs). Compared to the SMuRF model, MRFs are 2D models for which the dimensionality of the putative state-space is as large as the dimensionality of the raster, suggesting that MRFs may provide a more detailed characterizations of neural spike rasters. {Indeed, the SMuRF model makes the strong assumption that the neural spiking dynamics are decomposable into two time scales, with the additional simplifying assumption that there is only one component per time scale. This simplifying assumption is motivated by one-dimensional state-space models of neural data~\citep{smith2003estimating} in which a neuron's time-dependent CIF is only a function of \emph{one} hidden state sequence. We will investigate the inclusion of additional components in future work. We also plan to investigate analogues of the SMuRF model for population level data. MRFs, multi-component and population-level SMuRF models, naturally lead to model selection problems, and to the investigation of tools, based on sequential Monte-Carlo methods~\citep{chopin2013smc2} (aka particle filters), to compare state-space models of neural spike rasters (such as one-dimensional models~\citep{smith2003estimating}, the SMuRF model, and MRFs). The development of such tools for model comparisons is, in our opinion, the ultimate measure of the ability of different models to capture the intricate dynamics present in neural spike rasters}. Lastly, as previously mentioned, the SMuRF model can be interpreted as a two-dimensional Gaussian process prior on the neural spiking rate surface~\citep{rad2010efficient}, with a \emph{separable} kernel that is the Kronecker product of kernels from Gauss-Markov processes (one process for each dimension).  The choice of kernels in the SMuRF model leads to the very efficient algorithms for estimation and inference derived in this article. Moreover, these algorithms scale well to more than two dimensions unlike classical kernel methods. We plan to explore this connection to Gaussian process inference in future work.

\subsection*{Acknowledgements}
We would like to thank Dr. Anne C. Smith for her generous feedback on this manuscript and extensive discussions regarding the SMuRF model. Demba thanks the Alfred P. Sloan Foundation. K.M.T. is a New York Stem Cell Foundation--Robertson Investigator and McKnight Scholar and this work was supported by funding from the JPB Foundation, the PIIF and PIIF Engineering Award, PNDRF, JFDP, Alfred P Sloan Foundation, New York Stem Cell Foundation, McKnight Foundation, R01-MH102441-01 (NIMH), RF1-AG047661-01 (NIA), R01-AA023305-01 (NIAAA) and NIH Director’s New Innovator Award DP2-DK-102256-01 (NIDDK).

\section*{Appendix}

{{

\subsection*{The SMuRF model cannot be converted easily to a standard state-space model}
We focus on the simple case when $\rho_x = \rho_z = 1$, and $\alpha_x = \alpha_z = 0$
$$\left\{
\begin{array}{ll}
	  x_k =  x_{k-1} + \epsilon_k , \epsilon_k \sim \mathcal{N}(0, \sigma^2_\epsilon), k = 1,\cdots,K \\
      z_r =  z_{r-1} + \delta_r , \delta_r \sim \mathcal{N}(0,\sigma^2_\delta), r = 1,\cdots,R
\end{array}
\right. $$
\noindent Let $t = (r-1)\times K + k$, $r = 1, \cdots,R$, $k=1,\cdots,K$. The index $t$ is obtained by ``unstacking" the raster trials and serializing them.

The question we ask is whether the state equations from the SMuRF model can be turned into ones of the form
\begin{equation}
	\ess_t = \A \ess_{t-1} + \vv_t,
	\label{eq:ssm}
\end{equation}
\noindent where $\ess_t \in \R^2$. Let $\ess_{t,1}$ and $\ess_{t,2}$ denote the first and second components of $\ess_t$ respectively. We ask that $\ess_t \in \R^2$ because of the two dimensions present in the SMuRF model. Allowing the dimensionality of $\ess_t$ to increase up to $K$ would allow a representation of the form of Equation~\ref{eq:ssm}. However, this would become a very high-dimensional, unwieldy state-space model.

Intuitively, this cannot be done for the following reason: the dimensionality of the latent states in the SMuRF model is $K + R$, while the dimensionality of the state sequence in Equation~\ref{eq:ssm} is $2 \times (K \times R)$. For there to be an equivalence, the sequence $\ess_t$ must necessarily be redundant, i.e. some of the states must be copies of previous states. Storing these copies, would necessarily mean having to increase the dimensionality of the state space!

Let $\ess_t = \begin{bmatrix} x_{t - \left(\left\lceil \frac{t}{K} \right\rceil - 1\right) \times K  } \\ z_{\left\lceil \frac{t}{K} \right\rceil} \end{bmatrix} \in \R^2$. The quantity $\left\lceil \frac{t}{K} \right\rceil$ gives the trial index $r$ corresponding time index
$t$. The within-trial index corresponding to index $t$ is then obtained by substracting $(r-1) \times K$ from $t$.

Note, for instance, that $\ess_{1} = \begin{bmatrix} x_1 \\ r_1 \end{bmatrix}$ and $\ess_{K+1} = \begin{bmatrix} x_1 \\ r_2 \end{bmatrix}$. In general, $\ess_{t,1} = \ess_{t',1} = x_{k_0}$ for some $1 \leq k_0 \leq K$ if and only if $t > t'$  s.t. $t - t' = p\times K$ for some integer $p$, where we assume without loss of generality that $t > t'$. That is, two different indices $t$ and $t'$ share the same within-trial component if and only if they are apart by an integer multiple of $K$. Stated otherwise, the first component of $\ess_t$ exhibit circular symmetry! Therefore, for Equation~\ref{eq:ssm} to hold, $\ess_{t-1,1}$ must equal $\ess_{t,1}$, which is not possible because $t$ and $t-1$ are not apart by an integer multiple of $p$!

The argument above shows that, in order to write the SMuRF state equations in the form of Equation~\ref{eq:ssm}, one would need to augment the state $\ess_t$ to dimension $K + 1$, which would lead to a very high dimensional standard state-space model, thus increasing the complexity of performing inference.
}}

\subsection*{Derivation of Gibbs sampler for PG-augmented SMuRF model}
\noindent We first derive Theorem~\ref{theorem2}, which leads to the forward-filter backward-sampling algorithm from the full-conditionals for $\x$ and $\z$ in the Gibbs sampler.

Since $w_{k,r}| x_k,z_r$ is drawn from a PG distribution, we can write the log pdf of $w_{k,r}| x_k,z_r$ as,

\begin{align}
\log p(w_{k,r}| x_k,z_r) &= \log\left(\cosh\left(\frac{x_k+z_r}{2}\right)\right) + \log\left(\sum\limits_{i=1}^{\infty}(-1)^i \frac{(2i+1)}{\sqrt{2\pi {w_{k,r}}^2}} e^{-\frac{(2i+1)^2}{8w_{k,r}}-\frac{x_k^2 w_{k,r}}{2}}\right)
\end{align}

\noindent The complete data likelihood of the SMuRF model is,
\begin{align}
p(\mathbf{\Delta N}, \x, \w; \theta) &= p(\mathbf{\Delta N}|\x,\w)p(\w|\x)p(\x)\\
&= \prod\limits_{k=1}^{K}\prod\limits_{r=1}^{R}\left\{p(\Delta N_{k}^r|x_k,z_r)p(w_{k,r}|x_k,z_r)\right\}\prod\limits_{k=1}^{K} p(x_k|x_{k-1}; \sigma_{\epsilon}^2)\prod\limits_{r=1}^{R}p(z_r|z_{r-1}; \sigma_{\delta}^2)
\end{align}
The log of the complete data likelihood is therefore,
\begin{align}
&\log p(\mathbf{\Delta N}, \x, \w; \theta) \nonumber \\
&=  \sum\limits_{k=1}^{K} \sum\limits_{r=1}^{R} \left\{ \log p(\Delta N_{k,r}|x_k,z_r) + \log p(w_{k,r}|x_k,z_r)\right\} + \pi(\x,\z)\\
&= \sum\limits_{k=1}^{K} \sum\limits_{r=1}^{R} \left[\Delta N_{k,r}\log\left(\frac{e^{x_k+z_r}}{1+e^{x_k+z_r}}\right) + (1-\Delta N_{k,r})\log\left(\frac{1}{1+e^{x_k+z_r}}\right)+ \log\left(\cosh\left(\frac{x_k+z_r}{2}\right)\right)\right.\nonumber\\
&\,\,\,\,\,\, \left. + \log\left(\sum\limits_{i=1}^{\infty}(-1)^i \frac{(2i+1)}{\sqrt{2\pi {w_{k,r}}^2}} e^{-\frac{(2i+1)^2}{8w_{k,r}}-\frac{(x_k+z_r)^2w_{k,r}}{2}}\right)\right]+\pi(\x,\z)\\
&= \sum\limits_{k=1}^{K} \sum\limits_{r=1}^{R}\left[ \Delta N_{k,r}(x_k+z_r) - \log(1+e^{x_k+z_r}) + \log\left(\frac{1+e^{x_k+z_r}}{2e^{\frac{x_k+z_r}{2}}}\right) \right.\nonumber \\
&\,\,\,\,\,\, \left.+ \log \left(e^{-\frac{(x_k+z_r)^2w_{k,r}}{2}}\sum\limits_{i=1}^{\infty}(-1)^i \frac{(2i+1)}{\sqrt{2\pi {w_{k,r}}^2}} e^{-\frac{(2i+1)^2}{8w_{k,r}}}\right)\right]+\pi(\x,\z)\\
&= \sum\limits_{k=1}^{K}\sum\limits_{r=1}^{R}\left[ \Delta N_{k,r}(x_k+z_r) - \log(2)-\frac{x_k+z_r}{2} -\frac{(x_k+z_r)^2w_{k,r}}{2}\right.\nonumber\\
&\,\,\,\,\,\, \left.+ \log \left(\sum\limits_{i=1}^{\infty}(-1)^i \frac{(2i+1)}{\sqrt{2\pi {w_{k,r}}^2}} e^{-\frac{(2i+1)^2}{8w_{k,r}}}\right)\right]+\sum\limits_{k=1}^{K}\left[\frac{1}{2}\log(2\pi\sigma_{\epsilon}^2)-\frac{(x_k-x_{k-1})^2}{2\sigma_{\epsilon}^2}\right]+\pi(\x,\z)\\*
&=K\log(2)+ \pi(\x,\z) + \sum\limits_{k=1}^{K} \sum\limits_{r=1}^{R}{\left\{\Delta N_{k}^r(x_k+z_r)-\frac{x_k+z_r}{2}-\frac{(x_k+z_r)^2w_{k,r}}{2}\right\}}\nonumber\\
&\,\,\,\,\,\,+\sum\limits_{k=1}^{K}\sum\limits_{r=1}^{R}\left\{ \log \left(\sum\limits_{i=1}^{\infty}(-1)^i \frac{(2i+1)}{\sqrt{2\pi {w_{k,r}}^2}} e^{-\frac{(2i+1)^2}{8w_{k,r}}}\right)\right\}
\end{align}
where
\begin{align}
\pi(\x,\z) &= \sum\limits_{k=1}^{K}\left[\frac{1}{2}\log(2\pi\sigma_{\epsilon}^2)-{{\frac{(x_k-\rho_x x_{k-1} - \alpha_x u_{x,k})^2}{2\sigma_{\epsilon}^2}}}\right] \\ \nonumber
&\,\,\,\,\,\,+\sum\limits_{r=1}^{R}\left[\frac{1}{2}\log(2\pi\sigma_{\delta}^2)-{{\frac{(z_r-\rho_z z_{r-1}-\alpha_z u_{z,k})^2}{2\sigma_{\delta}^2}}}\right]
\end{align}

From the complete data log likelihood, we see that
\begin{align}
\log p(\x | \mathbf{\Delta N}, \z,\w;\theta) &\propto  \sum\limits_{k=1}^{K} \sum\limits_{r=1}^{R}{\left\{\Delta N_{k}^r(x_k+z_r)-\frac{x_k+z_r}{2}-\frac{(x_k+z_r)^2w_{k,r}}{2}\right\}} + \pi(\x,\z)\\
&\propto  -\sum\limits_{k=1}^{K} \sum\limits_{r=1}^{R}\frac{1}{2}\frac{(\Delta \tilde{N}_{k,r}-(x_k+x_r))^2}{1/w_{k,r}}+\pi(\x,\z),
\end{align}
where $\Delta \tilde{N}_{k,r} = \frac{\Delta N_{k,r} - \frac{1}{2}}{w_{k,r}}$.\\
Therefore, we can rewrite the augmented model as a linear Gaussian state space model as stated in Theorem \ref{theorem2}.
$$ \left\{\begin{array}{ll}
	{{x_k = \rho_x x_{k-1} + \alpha_x u_{x,k} + \epsilon_k}}, \epsilon_k \sim \mathcal{N}(0,\sigma^2_{\epsilon}) & \\
	\tilde{y}_k = x_k + \tilde{v_{k}}, \tilde{v_{k}} \sim \mathcal{N}\left(0,\left(\sum\limits_{r=1}^R w_{k,r}\right)^{-1}\right) &\\
       \tilde{y}_k = \Delta \tilde{N}_k = x_k-\frac{K}{2}-\sum\limits_{r=1}^R x_r w_{k,r}& \\
\end{array}
\right .$$
Let $H_{k,r} = {\Delta \tilde{N}_1^r,\dots,\Delta \tilde{N}_{k-1}^r}$ denote the history of the observed process up-to and including $k-1$. We can now write
\begin{align}
p(\x|\mathbf{\Delta N, \z, \w; \theta}) \propto\prod\limits_{k=1}^K {p(\Delta N|x_k)p(x_k|H_k)}
\end{align}
The forward filtering equations for this linear Gaussian state space model are as follows.
\begin{align}
\label{eqn:filter_xk1}
x_{k|k-1} &= {{\rho_x x_{k-1|k-1} + \alpha_x u_{x,k}}}\\
\label{eqn:filter_xk2}
\sigma^2_{k|k-1} &= {{\rho_x^2\sigma^2_{k-1|k-1}+\sigma^2_{\epsilon}}}
\end{align}
\begin{align}
\label{eqn:filter_xk3}
x_{k|k} &={{\rho_xx_{k|k-1}}} \\
&\,\,\,\,\,\, +{{\frac{\left(\sum\limits_{r=1}^Rw_{k,r}\right)\sigma^2_{k|k-1}}{1+\left(\sum\limits_{r=1}^Rw_{k,r}\right)\sigma^2_{k|k-1}}\left(\frac{\alpha_x u_{x,k}}{\left(\sum\limits_{r=1}^Rw_{k,r}\right)\sigma^2_{k|k-1}}+\frac{ \sum\limits_{r=1}^R \Delta N_{k}^r -\frac{R}{2}-\sum\limits_{r=1}^R z_r w_{k,r}}{\left(\sum\limits_{r=1}^Rw_{k,r}\right)}-\rho_x x_{k|k-1}\right)}}
\end{align}
\begin{align}
\label{eqn:filter_xk4}
\sigma^2_{k|k} &= {{\frac{\sigma^2_{k|k-1}}{1+{\left(\sum\limits_{r=1}^Rw_{k,r}\right)}\sigma^2_{k|k-1}}}}
\end{align}
After running the forward filtering algorithm, we obtain $x_{K|K}$ and $\sigma^2_{K|K}$ from the final iteration of the filter. We can then draw $x_K \sim N(x_{K|K}, \sigma^2_{K|K})$. Now we can treat $x_K$ as the new observations and use the Kalman filter again to draw samples for $x_{K-1}$, and repeat this process iteratively for $x_{K-1},..., x_1$. The new observation equation reads,
$$\left\{
\begin{array}{ll}
      x_k = {{\rho_x x_{k-1} + \epsilon_k}}, \epsilon_k \sim N(0,\sigma^2_{\epsilon}) \\
      x_{k+1} = x_{k}+\epsilon_{k}
\end{array}
\right. $$
From Bayes Rule we have,
\begin{align}
p(x_k|x_{k+1},H_k) = \frac{p(x_{k+1}|x_k)p(x_k|H_{k})}{p(x_{k+1}|H_{k})}
\end{align}
Denote the densities of $x_{k}|x_{k+1},H_k$ as
$$x_{k}|x_{k+1},H_k\sim N({x_{k|k}}^{*}, {\sigma^2_{k|k}}^{*})$$
Then the update equations are,
\begin{align}
\log p(x_k|H_k) &\propto \log p(x_{k+1}|x_k) + \log p(x_k|H_{k-1})\\
{x_{k|k}}^{*} &= {{{\rho_x x_{k|k-1} }+ \frac{{\sigma^2_{k|k}}}{\sigma^2_{\epsilon}}(x_{k+1}-{\rho_x x_{k|k-1}})}}
\end{align}
\begin{align}
{\sigma^2_{k|k} }^{*}&= {{\frac{\sigma^2_{\epsilon}{\sigma^2_{k|k-1}}}{\sigma^2_{\epsilon}+{\sigma^2_{k|k-1}}}}}
\end{align}
With this backward-sampling algorithm, we can draw $x_{k} \sim N({x_{k|k}}^{*},{\sigma^2_{k|k}}^{*})$, where $i = K-1, ..., 1$. The forward-filtering and backward-sampling algorithm are symmetric for $x_k$ and $z_r$.

\subsection*{Initialization of the EM algorithm and the Gibbs sampler}
We initialize the Monte-Carlo EM algorithm with values for $\sigma^2_{\epsilon}$ and $\sigma^2_{\delta}$ obtained by applying the one-dimensional state space model from \citep{smith2003estimating} to the raster data aggregated across either trials or time. We initialize the Gibbs sampler using trajectories drawn from posterior distribution of the state in the one-dimensional state-space models~\citep{smith2003estimating} used to initialize $\sigma^2_{\epsilon}$ and $\sigma^2_{\delta}$. The Gibbs sampler draws 5000 samples for $\x$, $\z$, and $\w$ at every iteration. The algorithm reaches convergence when the absolute change in $\sigma^2_\epsilon$ and $\sigma^2_\delta$ is less than a certain threshold ($10^{-5}$).

{

\subsection*{Ability of SMuRF model to identify learning time and trial in simulated data}

The results of our analysis of the cortical data in Section~\ref{sec:res} demonstrate that learning of a contingency by a neuron is a dynamic process that cannot be easily quantified in terms of a static time and trial of learning. We also demonstrated (Table~\ref{learning_time_trial_table}) how to use inferences from the SMuRF model to identify a learning time and a trial.

Here, we use simulated data to determine the ability of the SMuRF model to identify learning time and trial when learning is accompanied by \emph{sustained} changes in neural spiking following conditioned stimulus onset and during conditioning. In particular, we assess the sensitivity of our method to the extent of the change in neural spiking rate following conditioned stimulus onset and during conditioning.

We simulated neural spike raster data in the same manner as described in the \textbf{Simulation Studies} component of our \textbf{Applications} section (Section~\ref{sec:res}).  As in said section, the raster is divided into two regions (Figure~\ref{fig:simulated_region}). We assume that the rate of spiking of the neuron in Region B is fixed and equal $\lambda_B = 20$ Hz. We vary the rate of spiking $\lambda_A$ of the neuron in Region A from $20$ to $45$ Hz in $5$ Hz increments. For each value of $\lambda_A$, we simulated 10 independent rasters and determine the learning time and trial as in Table~\ref{learning_time_trial_table}. We use the average over the 10 rasters as the learning time/trial pair. When our method detects no change, we declare the learning time and trial as the last time and trail pair in the simulated data, i.e. $1000$ ms and trial. Figures~\ref{fig:avlrngtt}(a) and~\ref{fig:avlrngtt}(b) show the averages of the identified learning times and trials as a function of the ratio $\frac{\lambda_A}{\lambda_B}$. The true learning time is at $0$ ms with respect to conditioned stimulus onset, and true learning trial is trial $16$. The figures demonstrate that the inference performed from the SMuRF model is able to detect the true learning time and trial when the rate in Region A is $1.8$ and $2$ times larger than that in Region B. Moreover, the lower the ratio $\frac{\lambda_A}{\lambda_B}$, the larger the delay. The intuitive reason why it is easier to determine the learning trial is that, for a given trial, there are many more observations, compared to the number of trials for a give time instant.

\begin{figure}[H]
        \begin{center}
		\includegraphics[scale=1]{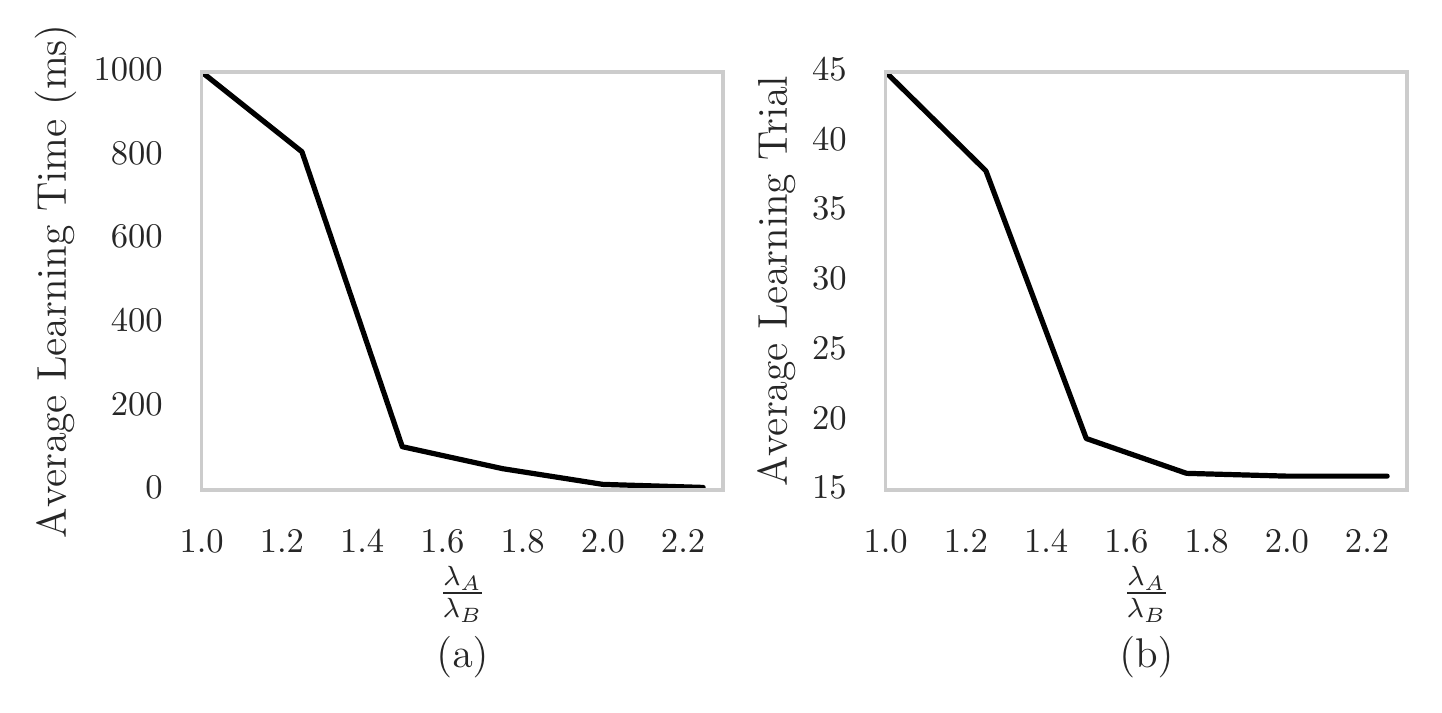}
        \end{center}

    \vspace*{-3mm}

    \caption{Plot of average (a) learning time and (b) learning trial identified by SMuRF inference as a function of ratio of neural spiking rate following and preceding learning. The learning time and trial are defined as the first time and trial pair when then empirical probability (Equation~\ref{eq:probev}) that the spiking rate at a given time/trial pair (Figure~\ref{fig:region}, Point C) is bigger than the average rate at the same trial (Figure~\ref{fig:region}, Region A) and the average rate at the same time (Figure~\ref{fig:region}, Region B) is larger than 95\%. The average is taken over 10 independently simulated rasters for each of the values for the ration $\frac{\lambda_A}{\lambda_B}$. The figure demonstrate the ability of inference performed using the SMuRF model to reliably identify learning.}

    \label{fig:avlrngtt}

\end{figure}

\subsection*{Robustness of SMuRF model to the presence of error trials}

We simulated neural spike raster data in the same manner as described in the \textbf{Simulation Studies} component of our \textbf{Applications} section (Section~\ref{sec:res}). We picked three consecutive trials, starting from trial $21$, to be a error trials in which all of the observations were $0$. Note that the data were simulated in the same manner as in Figure~\ref{fig:simulated_data}, except for the presence of the error trials. Figure~\ref{fig:raster_error_trial3} shows the result of applying the SMuRF model to these simulated raster data. The presence of the error trials does not affect our remarks for Figure~\ref{fig:simulated_data}.

\begin{figure}[H]
        \begin{center}
		\includegraphics[scale=1]{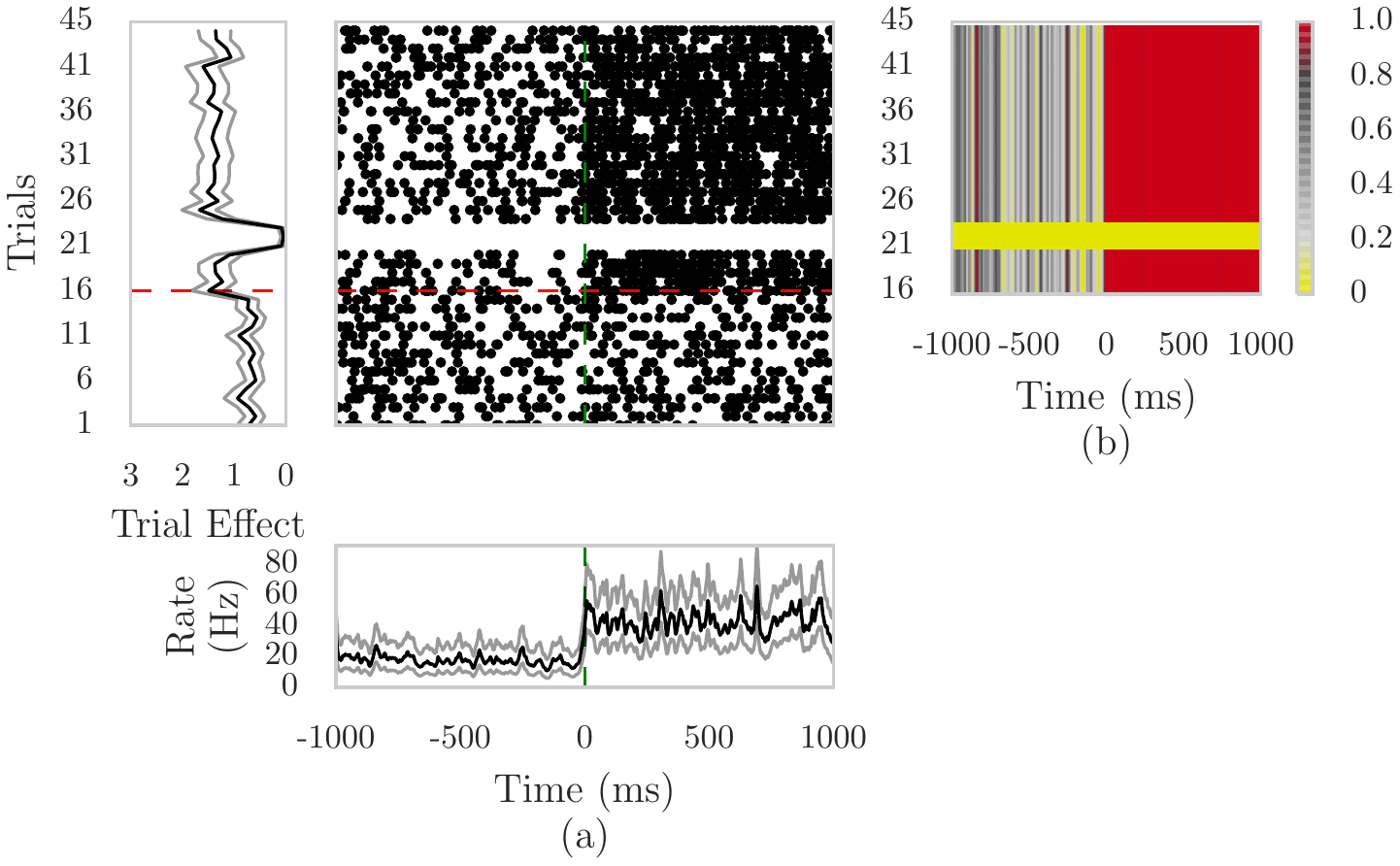}
        \end{center}

    \vspace*{-3mm}

    \caption{(a) Simulated neural spike raster with three consecutive error trials, along with estimated within and cross trial effects from the SMuRF model. The horizontal red line indicates the beginning of conditioning. The vertical green line indicates the onset of the conditioned stimulus. Despite the presence of the error trials, the left panel of the figure shows the cross-trial effect which, following conditioning, increases above its average initial value of $\approx 1$ to $\approx 2$.}

    \label{fig:raster_error_trial3}

\end{figure}
\addtocounter{figure}{-1}
\begin{figure} [h!]
    \captionof{figure}{{(b) Empirical probability that spiking rate at a given trial and time is bigger than the average rate at the same trial and the average rate during habituation at the same time. Despite the presence of the error trials, we can see that with probability close to $1$, the spiking rate at a given time/trial pair--following the conditioned stimulus and during conditioning (Figure~\ref{fig:region} C)--is bigger than the average rate at the same trial (Figure~\ref{fig:region} A) and the average rate at the same time (Figure~\ref{fig:region} B).}}
\end{figure}
}

%
%
%
%
%
%
%
%
%
%
%
%
%

\begin{figure}[H]
       \begin{center}

		\includegraphics[scale=1]{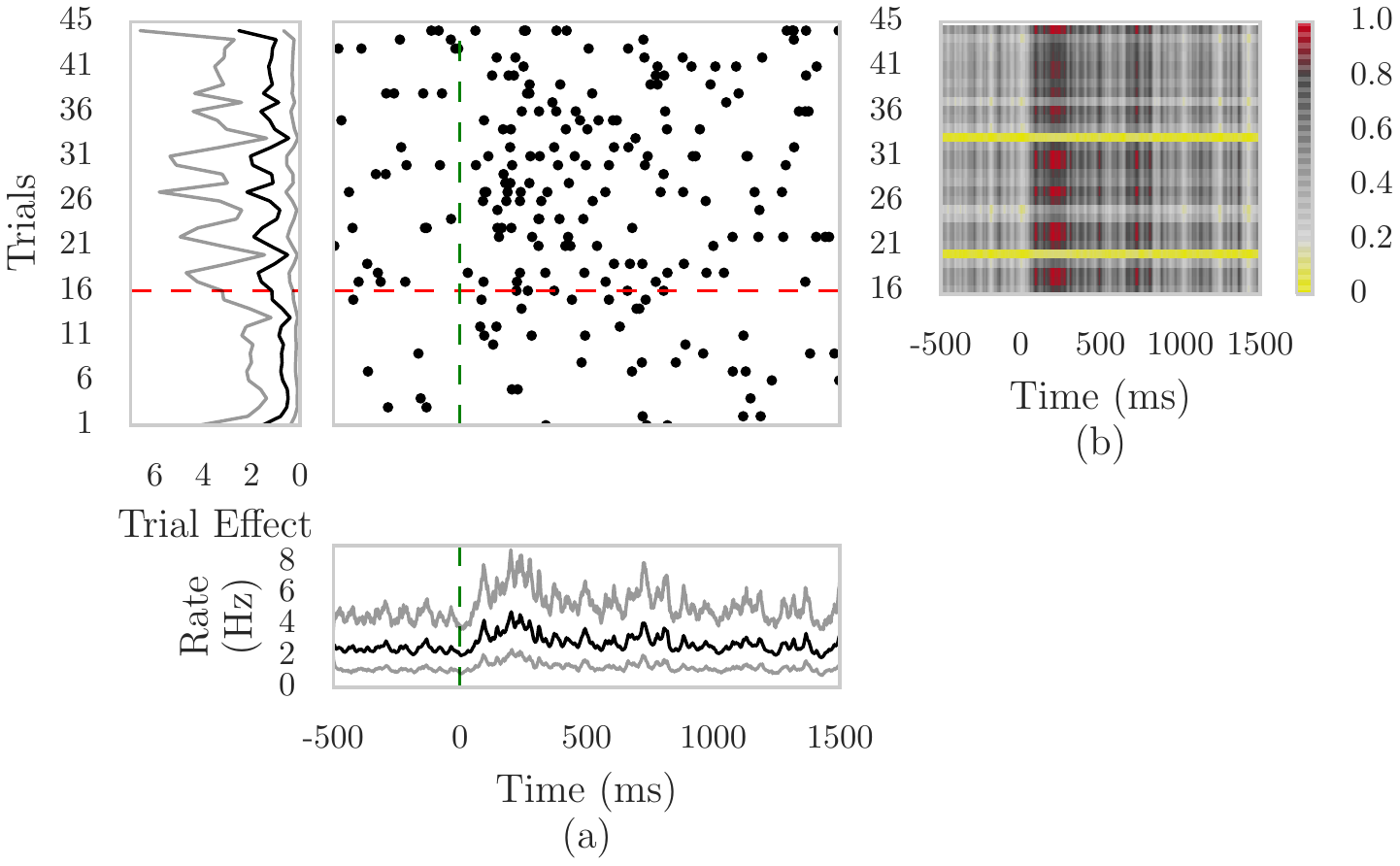}

        \end{center}

    \vspace*{-3mm}

  \caption{{(a) Raster, along with SMuRF within and cross-trial components, from a cortical neuron that exhibits a transient conditioned response to the conditioned stimulus. The horizontal red line indicates the beginning of conditioning. The vertical green line indicates the onset of the conditioned stimulus. The bottom panel suggests that this neuron exhibits a delayed response to the conditioned stimulus, beginning at $\approx < 100$ ms following conditioned stimulus presentation. The response then decreases and is followed by a slight increase at $\approx 700$ ms.}}

    \label{fig:new_raster_2193}

\end{figure}
\addtocounter{figure}{-1}
\begin{figure} [h!]
    \captionof{figure}{{Accounting for this increase in within-trial spiking rate due to the conditioned stimulus, the left panel shows a multiplicative increase in spiking rate due to conditioning from an average initial value of $\approx 1$ to a peak average value of $\approx 2$ between trial $ \approx 18$ at trial $30$. (b) Empirical probability that spiking rate at a given trial in and time is bigger than the average rate at the same trial and the average rate during habituation at the same time (refer to Figure~\ref{fig:region}). This panel provides a more detailed account of the the intricate dynamics in neural modulation that accompanies conditioning to the stimulus for this neuron.}}
\end{figure}
%
%
%
%
%
%
%
%
%
%
%
%
%
%
%
%

\begin{figure}[H]
       \begin{center}

		\includegraphics[scale=1]{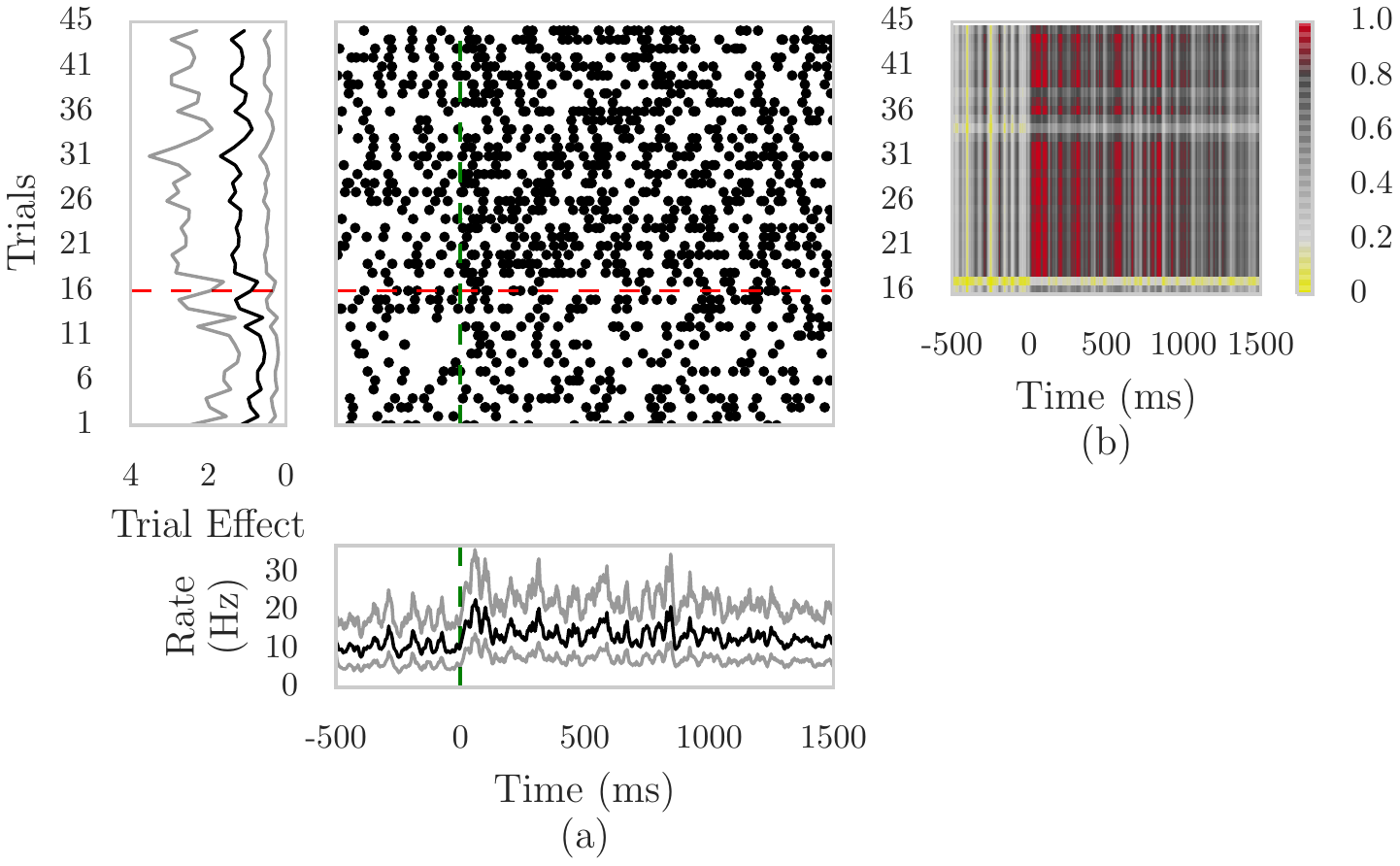}

        \end{center}

    \vspace*{-3mm}

  \caption{{(a) Raster, along with SMuRF within and cross-trial components, from a cortical neuron that exhibits a transient conditioned response to the conditioned stimulus. The horizontal red line indicates the beginning of conditioning. The vertical green line indicates the onset of the conditioned stimulus.}}

    \label{fig:new_raster_3189}

\end{figure}
\addtocounter{figure}{-1}
\begin{figure} [h!]
    \captionof{figure}{{The bottom panel suggests that this neuron exhibits a delayed response to the conditioned stimulus, beginning at $\approx 20$ ms following conditioned stimulus presentation. The response is sustained until $\approx 800$ ms, and then decreases. Accounting for this increase in within-trial spiking rate due to the conditioned stimulus, the left panel shows a multiplicative increase in spiking rate due to conditioning from an average initial value of $\approx 1$ to a peak average value of $\approx < 2$ between trial $ \approx 18$ at trial $30$. (b) Empirical probability that spiking rate at a given trial in and time is bigger than the average rate at the same trial and the average rate during habituation at the same time (refer to Figure~\ref{fig:region}). This panel provides a more detailed account of the the intricate dynamics in neural modulation that accompanies conditioning to the stimulus for this neuron.}}
\end{figure}
\bibliographystyle{apacite}
\bibliography{smurf_arxiv}

%
%
%

\end{document}